\documentclass[final,authoryear,3p,times]{elsarticle}
\usepackage{amssymb}
\usepackage{amsmath}

\journal{Journal of Theoretical Biology}

\begin{document}

\begin{frontmatter}

    \title{Electroencephalographic field influence on calcium momentum
    waves}

    \author[li]{Lester Ingber\corref{cor1}\fnref{label1}}
    \address[li]{Lester Ingber Research, Ashland, OR} \ead{ingber@alumni.caltech.edu}
    \ead[url]{http://www.ingber.com}
    \author[li]{Marco Pappalepore}
    \author[li]{Ronald R. Stesiak} \fntext[label1]{Corresponding Author}

    \begin{abstract}
        Macroscopic electroencephalographic (EEG) fields can be an
        explicit top-down neocortical mechanism that directly drives
        bottom-up processes that describe memory, attention, and other
        neuronal processes. The top-down mechanism considered is
        macrocolumnar EEG firings in neocortex, as described by a
        statistical mechanics of neocortical interactions (SMNI),
        developed as a magnetic vector potential $ \mathbf{A} $. The
        bottom-up process considered is $ \mathrm{Ca}^{2+} $ waves
        prominent in synaptic and extracellular processes that are
        considered to greatly influence neuronal firings. Here, the
        complimentary effects are considered, i.e., the influence of $
        \mathbf{A} $ on $ \mathrm{Ca}^{2+} $ momentum, $ \mathbf{p} $.
        The canonical momentum of a charged particle in an
        electromagnetic field, $ \mathbf{\Pi} = \mathbf{p} + q \mathbf{A}
        $ (SI units), is calculated, where the charge of $ \mathrm{Ca}^{2+}
        $ is $ q = - 2 e $, $ e $ is the magnitude of the charge of an
        electron. Calculations demonstrate that macroscopic EEG $
        \mathbf{A} $ can be quite influential on the momentum $ \mathbf{p}
        $ of $ \mathrm{Ca}^{2+} $ ions, in both classical and quantum
        mechanics. Molecular scales of $ \mathrm{Ca}^{2+} $ wave
        dynamics are coupled with $ \mathbf{A} $ fields developed at
        macroscopic regional scales measured by coherent neuronal firing
        activity measured by scalp EEG\@. The project has three main
        aspects:  fitting $ \mathbf{A} $ models to EEG data as reported
        here, building tripartite models to develop $ \mathbf{A} $
        models, and studying long coherence times of $ \mathrm{Ca}^{2+} $
        waves in the presence of $ \mathbf{A} $ due to coherent neuronal
        firings measured by scalp EEG\@. The SMNI model supports a
        mechanism wherein the $ \mathbf{p} + q \mathbf{A} $ interaction
        at tripartite synapses, via a dynamic centering mechanism (DCM)
        to control background synaptic activity, acts to maintain
        short-term memory (STM) during states of selective attention.
    \end{abstract}

    \begin{keyword}
        short-term memory \sep astrocytes \sep neocortical dynamics \sep
        vector potential

    \end{keyword}

\end{frontmatter}

\section{Introduction}
\subsection{Multiple scales}
There is a growing awareness of the importance of multiple scales in
many physical and biological systems, including neuroscience
\citep{Anastassiou+Perin+Markram+Koch2011,Nunez+Srinivasan+Ingber2013}.
As yet, there do not seem to be any explicit top-down mechanisms that
directly drive bottom-up processes that describe memory, attention, etc.
Of course, there are many top-down type studies demonstrating that
neuromodulator
\citep{Silberstein1995} and neuronal firing states, e.g., as defined by
electroencephalographic (EEG) frequencies, can modify the milieu of
individual synaptic and neuronal activity, which is still consistent
with ultimate bottom-up paradigms. However, there is a logical
difference between top-down milieu as conditioned by some prior external
or internal conditions, and some direct top-down processes that direct
cause bottom-up interactions specific to short-term memory (STM).

This study crosses molecular ($ \mathrm{Ca}^{2+} $ ions), microscopic (synaptic
and neuronal), mesoscopic (minicolumns and macrocolumns), and
macroscopic (regional scalp EEG) scales. Calculations support the
interaction between synchronous columnar firings large enough to be
measured by scalp EEG and molecular scales contributing to synaptic
activity: On one hand, the influence of macroscopic scales on molecular
scales is calculated via the evolution of $ \mathrm{Ca}^{2+} $ quantum
wave functions. On the other hand, the influence of $ \mathrm{Ca}^{2+} $
waves is described in the context of a statistical mechanics model that
already has been verified as calculating experimental observables,
aggregating and scaling up from synaptic activity, to columnar neuronal
firings, to regional synchronous activity fit to EEG while preserving an
audit trail back to underlying synaptic interactions.
\subsection{Magnetism influences in living systems}
An example of a direct physical mechanism that affects neuronal
processing not part of ``standard'' sensory influences is the
possibility of magnetic influences in birds at quantum levels of
interaction
\citep{Kominis2009,Rodgers+Hore2009,Solov'yov+Schulten2009}. It should be
noted that this is just a proposed mechanism
\citep{Johnsen+Lohmann2008}.

The strengths of magnetic fields in mammalian neocortex may be at a
threshold to directly influence synaptic interactions with astrocytes,
the largest number of cells among glia cells, as proposed for long-term
memory (LTM)
\citep{Gordon+Iremonger+Kantevari+Ellis-Davies+MacVicar+Bains2009} and
STM
\citep{Banaclocha2007,Pereira+Furlan2010,Pereira+Santos+Barros2013}.
Magnetic strengths associated by collective EEG activity at a columnar
level gives rise to even stronger magnetic fields. Columnar excitatory
and inhibitory processes largely take place in different neocortical
laminae, providing possibilities for more specific mechanisms.

Note that magnetic fields generated by axons, about $ 10^{-7} $T, are
generally small relative to the Earth's magnetic fields on the order of $
3\times 10^{-5} $~T (T $ = $ Tesla $ = $ kg-A$ ^{-1} $-s$ ^{-2} $). This
is often cited as a reason that magnetic fields due to firing neurons
are not influential in brain processes. However, this paper calculates
the oscillatory magnetic vector potential $ \mathbf{A} $ due to many
synchronous minicolumns from many macrocolumns, not the magnetic field
of a single axon or minicolumn. The effects of this oscillatory $
\mathbf{A} $, synchronized to $ \mathrm{Ca}^{2+} $ waves which
contribute to this process, are the issue.

More examples of quantum influences in ``wet'' noisy biological systems
have developed into more general paradigms stressing such phenomena
\citep{Huelga+Plenio2013}.
\subsection{SMNI context of $ \mathrm{Ca}^{2+} $ waves}
Since 1981, 30+ papers on a statistical mechanics of neocortical
interactions (SMNI) applied to columnar firing states, have detailed
properties of short-term memory --- e.g., capacity (auditory $ 7\pm 2 $
and visual $ 4\pm 2 $), duration, stability , primacy versus recency
rule, Hick's law --- and other properties of neocortex by scaling up to
macrocolumns across regions to fit EEG data
\citep{Ingber1982,Ingber1983,Ingber1984,Ingber1994,Ingber1997,Ingber2012a}.
SMNI not only details STM, testing SMNI at columnar scales, but also
scaled-SMNI at relatively macroscopic scales has very well modeled large
EEG databases, testing SMNI at regional scales.

Experimental research supports information/memory processing by coherent
columnar firings across many neurons
\citep{Liebe+Hoerzer+Logothetis+Rainer2012,Salazar+Dotson+Bressler+Gray2012}.
This experimental confirmation greatly enhances the importance the SMNI
approach. There most likely are additional neural mechanisms that
actually code information within the context of such synchronous firings
\citep{Kumar+Rotter+Aertsen2010,Stanley2013} and other interactions among
neurons and astrocytes, e.g., ephaptic coupling of cortical neurons
\citep{Quiroga+Fried+Koch2013} and additional electromagentic
interactions contributing to the extracellular medium in which these
interactions transpire
\citep{Buzsaki+Anastassiou+Koch2012}. The neuroscience community also is
accepting that long-term memories are not stored in individual neurons,
but in groups of neurons perhaps as small as macrocolumns
\citep{Quiroga+Fried+Koch2013}. In this latter context of graded
phase-locked non-firing contributions to electromagnetic fields, such
interactions may also contribute to $ \mathbf{A} $, which should be
studied in future projects. However, the magnitudes of current in
neocortex in this paper are taken from experimental data of active
synchronous firing states, not theoretical calculations, and so they
likely include much of the contribution from non-firing sources, albeit
not important details of how they affect firing states.

The influence on the momentum of a $ \mathrm{Ca}^{2+} $ ion from
macrocolumnar EEG fields as measured on the scalp, is thereby considered
as the processing of information. SMNI calculates the influence of these
regional synchronous firings at molecular scales that drive most
influential $ \mathrm{Ca}^{2+} $ waves across synapses. The roles of $
\mathrm{Ca}^{2+} $ in neocortex, while not completely understood, are
very well appreciated as being quite important in synaptic interactions,
e.g., in modulating the production of excitatory glutamic acid
\citep{Zorumski+Mennerick+Que1996}, albeit not in all synaptic
interactions
\citep{Adam-Vizi1992}. It also is possible that $ \mathrm{Ca}^{2+} $
waves may be instrumental in tripartite synaptic interactions of
astrocytes and neuronal synapses, in neocortex
\citep{Agulhon+Petravicz+McMullen+Sweger+Minton+Taves+Casper+Fiacco+McCarthy2008,Araque+Navarrete2010,Ross2012}
and hippocampus
\citep{Kuga+Sasaki+Takahara+Matsuki+Ikegaya2011}, although the role of
tripartite synapses in the adult brain has been disputed
\citep{Sun+McConnell+Pare+Xu+Chen+Peng+Lovatt+Han+Smith+Nedergaard2013}. $
\mathrm{Ca}^{2+} $ waves have been labeled ``glissandi'', and likely
mediate large-scale cerebral blood flow
\citep{Kuga+Sasaki+Takahara+Matsuki+Ikegaya2011}. Researchers are more
regularly examining glial cells to better understand neural processing
of information
\citep{Han+Chen+Wang+Windrem+Wang+Shanz+Xu+Oberheim+Bekar+Betstadt+Silva+Takano+Goldman+Nedergaard2013}.
The $ \mathrm{Ca}^{2+} $ waves considered here arise from a nonlinear
cooperative regenerative process from internal stores, complementary to
and sometimes combining with $ \mathrm{Ca}^{2+} $ released through
classic endoplasmic reticulum channels and voltage-gated and
ligand-gated $ \mathrm{Ca}^{2+} $ transients. This likely includes a
process wherein $ \mathrm{Ca}^{2+} $ released from an inositol
triphosphate receptor (IP$ _{3} $R), requiring the presence of IP$ _{3} $,
acts on the same or other IP$ _{3} $R to release more $ \mathrm{Ca}^{2+}
$ while IP$ _{3} $ is still present. This process requires or affects
additional coupled processes involving other $ \mathrm{Ca}^{2+} $,
metabotropic glutamate receptors (mGluR), muscarinic acetycholine
receptors (mAChR), etc
\citep{Ross2012}.

Columnar EEG firings calculated by SMNI lead to electromagnetic fields
which can be described by a vector potential. This is referred to as the
SMNI vector potential (SMNI-VP). An early discussion of SMNI-VP
contained in a review of short-term memory as calculated by SMNI was
simply suggested
\citep{Ingber2012a}, and a previous paper outlined the approach taken
here, but only in a classical physics context
\citep{Ingber2012b}. Current research is directed to more detailed
interactions of SMNI-VP firing states with $ \mathrm{Ca}^{2+} $ waves.
\subsection{Outline}
Section 2 calculates the influence of $ \mathbf{A} $, derived using
current experimental data, on $ \mathrm{Ca}^{2+} $ momenta $ \mathbf{p} $
in both classical and quantum physics. Since $ \mathrm{Ca}^{2+} $ waves
are influential in synaptic interactions, this demonstrates the
influence of $ \mathbf{A} $ in synaptic interactions.

In Section 3 SMNI is scaled to regional EEG activity, e.g., as measured
on the scalp, and similar to previous studies that fit synchronous
columnar firings to the electric potential $ \Phi $ to EEG data, here
such synchronous columnar firings are scaled to the vector potential $
\mathbf{A} $. In either case, developing $ \Phi $ or $ \mathbf{A} $,
SMNI provides an audit trail back to columnar averaged synaptic
parameters which are fit to EEG data, with parameters constrained to
their experimentally determined ranges. In fitting $ \mathbf{A} $, the
prior established dependence of synaptic interactions on $ \mathbf{A} $
provides justification to include coefficients as parameters to test and
detail the dependence and sensitivity of EEG on $ \mathbf{A} $.

Section 4 discusses three current projects in the context of this paper.
Some early results are given for one of these projects, setting up the
framework for fitting SMNI models of synchronous regional firings to EEG
data. This framework includes dependence on models of synaptic
background activity, in turn dependent on $ \mathrm{Ca}^{2+} $ wave
influences from tripartite interactions, thereby providing some tests of
these multiple-scale models. A relatively large subsection details how
the SMNI model supports a mechanism wherein the $ \mathbf{p}+q\mathbf{A}
$ interaction at tripartite synapses, via a dynamic centering mechanism
(DCM) to control background synaptic activity, acts to maintain STM
during states of selective attention.

The first author (LI) is responsible for this main text, the computer
codes, and graphs of EEG data and CMI data as presented here. The other
authors (RS and MP) are responsible for their Supplementary analysis
summarized in Section 4.1.5.

The conclusion, Section 5, summarizes the calculations, which encourages
further investigations.
\section{Classical and quantum considerations}
In these calculations, the Lagrangian formulation will be used. In
descriptive terms, for classical physics calculations the Lagrangian $ L
$ is defined by the argument of a short-time conditional probability
distribution $ P $ over a vector of variables $ x $ and time $ t $,

\begin{equation}
    P[x(t)|x(t-\Delta t)]=\bar{\mathrm{N}}\exp (-L\Delta t)
\end{equation}
where $ \bar{\mathrm{N}} $ is a normalization prefactor. This
conditional probability evolves the initial distribution, e.g., as
expressed by the path integral over all variables at all intermediate
times. In quantum physics the Lagrangian is similarly defined in terms
of the evolution of the wave function $ \psi $ whose absolute square is
a probability distribution. As demonstrated in many disciplines as well
as in many SMNI papers, the Lagrangian formulation often offers
intuitive, algebraic and numerical advantages to its equivalent partial
and stochastic differential representations. E.g., this approach affords
the use of powerful derivations based on the associated variational
principle, e.g., Canonical Momenta and Euler-Lagrange equations. This is
all rigorously derived in many preceding SMNI papers, and has required
developing powerful numerical algorithms to fit these algebraic models
to data, such as Adaptive Simulated Annealing (ASA)
\citep{Ingber1993,Ingber2012c}, and to calculate numerical details of the
propagating probability distributions using PATHINT
\citep{Ingber+Nunez1995} and PATHTREE
\citep{Ingber+Chen+Mondescu+Muzzall+Renedo2001}.
\subsection{Effective momentum $ \mathbf{\Pi } $}
The effective momentum, $ \mathbf{\Pi } $, affecting the momentum $
\mathbf{p} $ of a moving particle in an electromagnetic field, is
understood from the canonical momentum
\citep{Feynman1961,Feynman+Leighton+Sands1964,Goldstein1980,Semon+Taylor1996},
in SI units,

\begin{equation}
    \mathbf{\Pi }=\mathbf{p}+q\mathbf{A}
\end{equation}
where $ q=-2e $ for $ \mathrm{Ca}^{2+} $, $ e $ is the magnitude of the
charge of an electron $ =1.6\times 10^{-19} $~C (Coulomb), and $ \mathbf
{A} $ is the electromagnetic vector potential. (In Gaussian units $
\mathbf{\Pi }=\mathbf{p}+q\mathbf{A}/c $, where $ c $ is the speed of
light.) $ \mathbf{A} $ represents three components of a 4-vector
\citep{Jackson1962,Semon+Taylor1996}. In the standard gauge, the 3-vector
components of this 4-vector potential related to magnetic fields are of
interest.

$ \mathbf{\Pi } $ can be used in quantum as well as in classical
calculations. Quantum mechanical calculations including these effects
are likely important as it is clear that in time scales much shorter
than neuronal firings $ \mathrm{Ca}^{2+} $ wave packets spread over
distances the size of typical synapses
\citep{Stapp1993}. The gauge of $ \mathbf{A} $ is not specified here, and
this can lead to important effects especially at quantum scales
\citep{Tollaksen+Aharonov+Casher+Kaufherr+Nussinov2010}.
\subsection{Quantum calculation}
The Lagrangian $ L $, the argument of the exponential defining this
probability distribution, includes the canonical energy $ \mathbf{\Pi }^
{2}/(2m) $. The momentum representation of a Gaussian wave function is
developed in this context. The magnetic vector potential field $ \mathbf
{A} $ is shown to be quite insensitive to a reasonable spatial location,
so we just have to consider the expectation of momentum p, which
essentially gives back the classical value. This is made more explicit
as follows:

The three-dimensional Gaussian wave function in momentum $ \mathbf{p} $-space
of a $ \mathrm{Ca}^{2+} $ ion is derived as follows. A wave packet of
many ions could be modeled by scaling the mass $ m $. The normalized
wave function at time $ t=0 $ in momentum space for a wave packet
centered with momentum $ \mathbf{p} $ is

\begin{equation}
    \phi (\mathbf{p},0)=(2\pi (\Delta \mathbf{p})^{2})^{-3/4}e^{-(\mathbf
    {p}-\mathbf{p}_{0})^{2}/(4(\Delta \mathbf{p})^{2})}
\end{equation}
where squared vectors represent inner products, e.g., $ (\Delta \mathbf{p})^
{2}=\Delta \mathbf{p} \cdot \Delta \mathbf{p} $. There is as yet no
experimental evidence as to how this kind of wave packet is developed by
$ \mathrm{Ca}^{2+} $ waves in vivo. $ \phi $ develops in time as $ U=\exp
(-iHt) $ with Hamiltonian/Energy $ H $,

\begin{equation*}
    U(\mathbf{p},t)=e^{-i((\mathbf{p}+q\mathbf{A})^{2}t)/(2m \hbar )}
\end{equation*}

\begin{equation}
    \phi (\mathbf{p},t)=\phi (p,0)U(p,t)
\end{equation}
Future analysis of an initial many-body wave function for a $ \mathrm{Ca}^
{2+} $ wave can be formulated as such a packet, and then developed in
time with intermittent collisions among constituents.

The normalized wave function in coordinate $ \mathbf{r} $-space is given
by a Fourier transform in $ \mathbf{k} $-space, which can be taken in $
\mathbf{p} $-space using $ \mathbf{p}= \hbar \mathbf{k} $,

\begin{equation}
    \psi (\mathbf{r},t)=(2\pi  \hbar )^{-3/2}\int \limits_{-\infty }\limits^
    {\infty }d^{3}\mathbf{p}\phi (p,t)e^{i\mathbf{p} \cdot \mathbf{r}/ \hbar }
\end{equation}
This integral yields

\begin{equation*}
    \psi (\mathbf{r},t)=\alpha ^{-1}e^{-\beta /\gamma -\delta }
\end{equation*}

\begin{equation*}
    \alpha =(2 \hbar )^{3/2}(2\pi (\Delta \mathbf{p})^{2})^{3/4}\left (\frac
    {it}{2m \hbar }-\frac{1}{4(\Delta \mathbf{p})^{2}}\right )^{3/2}
\end{equation*}

\begin{equation*}
    \beta =\left (\mathbf{r}-\frac{q\mathbf{A}t}{m}-\frac{i \hbar \mathbf{p}_
    {0}}{2(\Delta \mathbf{p})^{2}}\right )^{2}
\end{equation*}

\begin{equation*}
    \gamma =4\left (\frac{it \hbar }{2m}+\frac{\hbar ^{2}}{4(\Delta \mathbf
    {p})^{2}}\right )
\end{equation*}

\begin{equation}
    \delta =\frac{\mathbf{p}_{0}^{2}}{4(\Delta \mathbf{p})^{2}}+\frac{iq^
    {2}\mathbf{A}^{2}t}{2m \hbar }
\end{equation}
In coordinate space $ \psi $ exhibits a direct dependence on $ \mathbf{A}
$ in the displacement $ \mathbf{r}\rightarrow \mathbf{r}-q\mathbf{A}t/m $.
The phase-dependence of $ \mathbf{A} $, recognized as the Aharonov-Bohm
effect
\citep{Aharonov1959}, is not the issue here.

Note that

\begin{equation}
    (\Delta \mathbf{p})^{2}(\Delta \mathbf{r})^{2} \geq ( \hbar /2)^{2}
\end{equation}
where $ ((\Delta \mathbf{r})^{2})^{1/2} $ is the spatial 1/2-width of
the packet. With the variance of $ \psi $ in terms of $ 1/\Delta \mathbf
{p} $ instead of $ \Delta \mathbf{r} $, a factor of $  \hbar ^{-3/2} $ is
introduced into $ \psi (\mathbf{r},t) $ in order that the wave function
in coordinate space be properly normalized. The dispersion of the wave
packet in time can be seen in the factors and terms in $ \psi (\mathbf{r},t)
$, $ \{ \alpha ,\beta ,\gamma ,\delta \} $, above.

If just the effects of $ \mathbf{A} $ on the wave function is required,
using $ \mathbf{p} $-space is more straightforward than a typical $
\mathbf{p} \cdot \mathbf{A} $ calculation that does a partial
integration to get $ \partial \mathbf{A}/\partial t $, giving $ -\mathbf
{r} \cdot \mathbf{E} $, in terms of the coordinate $ \mathbf{r} $ and
electric field $ \mathbf{E} $, but $ \mathbf{r} $ is not as directly
observed as is $ \mathbf{p} $. Also note that the quantum expected value
of $ \mathbf{p} $ from $ \int \phi ^{*}\phi \mathbf{p} $ returns just $
\mathbf{p}_{0} $, the same as the classical value.
\subsection{$ \mathbf{A} $ of wire}
For a wire/neuron carrying a current $ \mathbf{I} $, measured in A (not
bold $ \mathbf{A} $) $ = $ Amperes $ = $ C/s,

\begin{equation}
    \mathbf{A}(t)=\frac{\mu }{4\pi }\int \frac{dr}{r}\mathbf{I}
\end{equation}
where the current is along a length $ z $ (a neuron), observed from a
perpendicular distance $ r $ from the line of thickness $ r_{0} $.
Neglecting far-field retardation effects, this yields

\begin{equation}
    \mathbf{A}=\frac{\mu }{4\pi }\mathbf{I}\log \left (\frac{r}{r_{0}}\right
    )
\end{equation}
Similar formulae for other geometries are in texts
\citep{Jackson1962}. The point here is the insensitive log dependence on
distance. The estimates below assume this log factor to be of order 1.
However, especially in this neocortical EEG context, the (oscillatory)
time dependence of $ \mathbf{A}(t) $ derived from $ \mathbf{I}(t) $ is
influential in the dynamics of $ \mathrm{Ca}^{2+} $ waves.

The magnetic field $ \mathbf{B} $ derived from $ \mathbf{A} $,

\begin{equation}
    \mathbf{B=\nabla \times A}
\end{equation}
is still attenuated in the glial areas where $ \mathrm{Ca}^{2+} $ waves
exist, and its magnitude decreases as inverse distance, but $ \mathbf{A}
$ derived near the minicolumns will be used there and at further
distances since it is not so attenuated. The electrical dipole for
collective minicolumnar EEG derived from $ \mathbf{A} $ is

\begin{equation}
    \mathbf{E}=\frac{ic}{\omega }\mathbf{\nabla \times B}=\frac{ic}{\omega
    }\mathbf{\nabla \times \nabla \times A}
\end{equation}
$ \mu _{0} $, the magnetic permeability in vacuum $ =4\pi 10^{-7} $ H/m
(Henry/meter), where Henry has units of kg-m-C$ ^{-2} $, is the
conversion factor from electrical to mechanical variables. Here $ \omega
$ represents a field with a single frequency, whereas if $ \mathbf{E} $
were the subject of study a dispersion relation would be required; this
gives context to the nature of $ \mathbf{A} $. In neocortex, $ \mu
\approx \mu _{0} $
\citep{Biswas+Luu2011,Georgiev2003}.

The contribution to $ \mathbf{A} $ can be viewed as including many such
minicolumnar lines of current across 100's to 1000's of macrocolumns
that typically contribute to large synchronous bursts of EEG
\citep{Srinivasan+Winter+Ding+Nunez2007}, e.g., within a region not large
enough to include many convolutions. Note that $ \mathbf{E} $ and $
\mathbf{B} $ do not possess the logarithmic insensitivity to distance
from active minicolumns as does $ \mathbf{A} $, and therefore they do
not possess the same advantage of approximately linearly accumulating
strength within macrocolumns.
\subsection{Effects of $ \mathbf{A} $ on $ \mathbf{p} $}
The momentum $ \mathbf{p} $ at issue is calculated for comparison to the
vector potential. In neocortex, a $ \mathrm{Ca}^{2+} $ ion with mass $
m=6.6\times 10^{-26} $~kg, has speed on the order of 50~$ \mu $m/s
\citep{Bellinger2005} to 100~$ \mu $m/s
\citep{Kuga+Sasaki+Takahara+Matsuki+Ikegaya2011,Ross2012}. This gives a
momentum on the order of $ 10^{-30} $~kg-m/s. A wave of many ions could
be modeled by scaling the mass $ m $. A study of molar concentrations
gives an estimate of a $ \mathrm{Ca}^{2+} $ wave as comprised of tens of
thousands of free ions representing about 1\% of a released set (the
bulk being buffered), with a range of about 100~$ \mu $m, sometimes as
much as 250~$ \mu $m
\citep{Bowser+Khakh2007}, duration of more than 500~ms, and
concentrations [$ \mathrm{Ca}^{2+} $] ranging from 0.1-5~$ \mu $M ($ \mu
$M = $ 10^{-3} $~mol/m$ ^{3} $)
\citep{Ross2012}.

$ q\mathbf{A} $ can be calculated at several scales:

In studies of small ensembles of neurons
\citep{Murakami+Okada2006}, an electric dipole moment $ \mathbf{Q} $ is
defined as $ \mathbf{I}z $ where $ z $ is in the direction of $ \mathbf{I}
$, leading to estimates of $ |\mathbf{Q}| $ for a pyramidal neuron on
the order of 1~pA-m = $ 10^{-12} $~A-m. Multiplying by $ 10^{4} $
synchronous firings in a macrocolumn gives an effective dipole moment $
|\mathbf{Q}|=10^{-8} $~A-m. Taking $ z $ to be $ 10^{2}\mu $m $ =10^{-4}
$~m (a couple of neocortical layers) to get $ \mathbf{I} $, this gives
an estimate $ |q\mathbf{A}|\approx 2\times 10^{-19}\times 10^{-7}\times
10^{-8}/10^{-4} $ = $ 10^{-28} $~kg-m/s,

Estimates at larger scales
\citep{Nunez+Srinivasan2006} give a dipole density $ |\mathbf{P}|= $ 0.1~$
\mu $A/mm$ ^{2} $. Multiplying this density by a volume of $ \mathrm{mm}^
{2}\times 10^{2}\mu \mathrm{m} $ (using the same estimate above for $ z $),
gives a $ |\mathbf{Q}|=10^{-9} $~A-m. This is smaller than that above,
due to this estimate including cancellations giving rise to scalp EEG,
while the estimate above is within a macrocolumn (the focus of this
study), leading to $ |q\mathbf{A}|=10^{-29} $~kg-m/s.

The estimates for $ \mathbf{Q} $ come from experimental data, which
therefore include all shielding and material effects expected in other
theoretical treatments that would derive $ \mathbf{Q} $. In the context
of coherent activity among many macrocolumns, correlated with STM
\citep{Salazar+Dotson+Bressler+Gray2012}, $ |\mathbf{A}| $ may become
orders of magnitude larger than these conservative estimates. Since $
\mathrm{Ca}^{2+} $ waves play an important role in synaptic activity
inherent in this coherent macrocolumnar activity, of course there is
direct coherence between these waves and the activity of $ \mathbf{A} $.

In the context of classical physics, the above calculations show how
important the effect of $ q\mathbf{A} $ from macroscopic EEG, on the
order of $ 10^{-28} $~kg-m/s can be on the momentum $ \mathbf{p} $ of a $
\mathrm{Ca}^{2+} $ ion, on the order of $ 10^{-30} $~kg-m/s. By itself,
this simple numerical comparison shows the important influence of $
\mathbf{A} $ on $ \mathbf{p} $ at classical scales.

In the context of quantum physics, the EEG effect on the displacement of
the $ \mathbf{r} $ coordinate in the $ \psi $ wave function, $ qAt/m $,
is on the order of $ 1.5\times 10^{-2}t $~m which within 100~ms is on
the order of $ 1.5\times 10^{-3} $~m. If we assume the extent of $
\Delta \mathbf{r} $ can be on the order of a synapse
\citep{Stapp1993}, then this spatial extent is on the order of about $
\mu $m $ =10^{4} $~$ \AA $ ($ \AA= $ Angstrom $ =10^{-10} $~m). (Typical
synaptic gaps are on the order of a few nm.) If this is correct, then
the displacement of $ \mathbf{r} $ by the $ \mathbf{A} $ term is much
larger than $ \Delta \mathbf{r} $. If the uncertainty principle is close
to saturation, we can take $ \Delta \mathbf{p} \geq  \hbar (2\Delta \mathbf{r})
$ = $ 1.054\times 10^{-34}/(2\times 10^{-6})=5\times 10-29 $~kg-m/s.
This would make $ \Delta \mathbf{p} $ about the same as $ \mathbf{p} $.
Given this spread for most ions in a wave, it is reasonable to further
investigate this ``beam'' of ions with respect to their entanglement.
\subsection{Quantum coherence of $ \mathrm{Ca}^{2+} $ waves}
While $ \mathrm{Ca}^{2+} $ are observed to remain in waves for durations
up to 500~ms
\citep{Ross2012}, the example above invokes extremely long quantum
coherence times of 100~ms, and even considerations on how long quantum
coherence times may be achieved may not support these long times. In any
case, it should be noted that $ \mathbf{A} $ exerts strong quantum
influences on $ \mathbf{r} $ via its relative influence on $ \mathbf{p} $.
So, while the $ \mathbf{p} $ wave packet may not survive for long
coherence times, the essential point is that there is experimental
verification of relatively free $ \mathrm{Ca}^{2+} $ ions surviving for
hundreds of ms, and these ions will be continuously affected by $
\mathbf{A} $. The above calculations of the effects of $ \mathbf{A} $ on
$ \mathbf{p} $ wave packets at least demonstrate the nature of this
interaction.

However, there are reasons to consider effects that may promote long
quantum coherence times. It is now understood that standard arguments,
that quantum coherence cannot be maintained at high temperatures
\citep{Davies2004}, simply may not apply to many complex biological
systems where other interactions may take precedence
\citep{Aharony+Gurvitz+Tokura+Entin-Wohlman+Dattagupta2012,Chin+Prior+Rosenbach+Caycedo-Soler+Huelga+Plenio2013,Fleming+Huelga+Plenio2011,Hartmann+Dur+Briegel2006,Lloyd2011}.
Quantum coherence in potassium ion channels has been proposed
\citep{Vaziri+Plenio2010}. A model that has some overlap with the present
context of the cooperative regenerative process that develops $ \mathrm{Ca}^
{2+} $ waves studies a free ion passing over two bound charges via
Coulomb interactions, leading to extended entanglement of the bound
charges as well as mediating extended entanglement with free ions
\citep{Buscemi+Bordone+Bertoni2007,Buscemi+Bordone+Bertoni2011}. In this
context, multiple sources of $ \mathrm{Ca}^{2+} $ ions that contribute
to a wave may develop enhanced entanglement among a significant number
of free ions.

Here, waves of free $ \mathrm{Ca}^{2+} $ ions
\citep{Ross2012}, the ions being synchronized into coherent waves by
phase coordination in this columnar coherent firing context
\citep{Pereira+Furlan2009}, may introduce pulsed-dynamical decoupling, a
generalization of the quantum Zeno effect (QZE) and ``bang-bang'' (BB)
decoupling, of ions from their environment, promoting long coherence
times
\citep{Facchi+Lidar+Pascazio2004,Facchi+Pascazio2008,Wu+Wang+Yi2012}, as
the system receives $ n $ ``kicks'' during time $ t $,

\begin{equation}
    U_{n}(p,t)=[U_{k}U(p,t/n)]^{n}
\end{equation}
where the kicks $ U_{k} $ may include interactions with other quantum
systems, e.g., other $ \mathrm{Ca}^{2+} $ ions in the same wave produced
in the regenerative processes discussed previously. Such mechanisms for
maintaining coherence are currently investigated in the context of
quantum computation
\citep{Rego+Santos+Batista2009,Yu+ZaiRong+Wei2012}. Distinguishable
particles, sometimes even if previously uncorrelated, can exhibit
quantum coherence and entanglement via collisions
\citep{Benedict+Kovacs+Czirjak2012,Harshman+Singh2008}, e.g., such as
collisions via Coulomb interactions among ions being synchronously
influenced by $ \mathbf{A} $ in $ \mathrm{Ca}^{2+} $ waves. Some
examples demonstrate how environment noise may favor extended
entanglement of quantum states
\citep{Zhang+Fan2013}.

In the context of of quantum $ \mathrm{Ca}^{2+} $ waves interacting with
$ \mathbf{A} $, while it is straightforward to model many-body
Hamiltonians/Lagrangians and wave functions, this issue can only be
resolved by experimental verification, e.g., to ascertain the degree of
quantum coherence among ions in a $ \mathrm{Ca}^{2+} $ wave. There is as
yet no experimental evidence as to how this long-time coherence is
developed by $ \mathrm{Ca}^{2+} $ waves in vivo.
\section{Coupled SMNI-VP $ \mathrm{Ca}^{2+} $-waves}
\subsection{SMNI dipoles}
A dipole model for collective minicolumnar oscillatory currents is
considered, corresponding to top-down signaling, flowing in ensembles of
axons, not for individual neurons. The top-down signal is claimed to
cause relevant effects on the surrounding milieu, but is not appropriate
outside these surfaces due to strong attenuation of electrical activity.
However, the vector potentials produced by these dipoles due to axonal
discharges do survive far from the axons, and this can lead to important
effects at the molecular scale, e.g., in the environment of ions
\citep{Feynman+Leighton+Sands1964,Giuliani2010}.

The SMNI columnar probability distributions, derived from statistical
aggregation of synaptic and neuronal interactions among minicolumns and
macrocolumns, have established credibility at columnar scales by
detailed calculations of properties of STM\@. Under conditions enhancing
multiple attractors, detailed in SMNI with a ``centering mechanism'' (CM)
effected by changes in background synaptic activity, multiple columnar
collective firing states are developed. These minicolumns are the
entities which the above dipole moment is modeling. The Lagrangian of
the SMNI distributions, although possessing multivariate nonlinear means
and covariance, has a functional form similar to arguments of firing
distributions of individual neurons, so that the description of the
columnar dipole above is a model faithful to the standard derivation of
a vector potential from an oscillating electric dipole.

Note that this is not necessarily the only or most popular description
of electromagnetic influences in neocortex, which often describes
dendritic presynaptic activity as inducing large scale EEG
\citep{Nunez1981}, or axonal firings directly affecting astrocyte
processes
\citep{McFadden2007}. This work is only and specifically concerned with
electromagnetic fields in collective axonal firings, directly associated
with columnar STM phenomena in SMNI calculations, which create vector
potentials influencing ion momenta just outside minicolumnar structures.
\subsection{SMNI Lagrangian}
A very short summary of the relevant SMNI Lagrangian in terms of its
scaled synaptic parameters enables an explicit presentation of coupling
the SMNI-VP with $ \mathrm{Ca}^{2+} $ waves.

Care was taken in the first derivations of SMNI to properly process
time-dependent and nonlinear multivariate drifts and diffusions. E.g.,
in the mid-point (Stratonovich or Feynman) representation, all
Riemannian contributions were calculated and numerically estimated for
neocortex, as the nonlinear multivariate diffusions present a curved
space
\citep{Ingber1982,Ingber1983}. A derivation of the underlying
mathematical physics has been in some specialized text books for some
time
\citep{Langouche+Roekaerts+Tirapegui1982}, and a compact derivation has
been given in several papers
\citep{Ingber1991}.

The SMNI Lagrangian, $ L $, in the prepoint (Ito) representation was
derived as

\begin{equation*}
    L=\sum \limits_{G,G^\prime{}}(2N)^{-1}(\dot{M}^{G}-g^{G})g_{GG^\prime
    {}}(\dot{M}^{G^\prime{}}-g^{G^\prime{}})/(2N\tau )-V^\prime{}
\end{equation*}

\begin{equation*}
    g^{G}=-\tau ^{-1}(M^{G}+N^{G}\tanh F^{G})
\end{equation*}

\begin{equation*}
    g^{GG^\prime{}}=(g_{GG^\prime{}})^{-1}=\delta _{G}^{G^\prime{}}\tau ^
    {-1}N^{G}\mathrm{sech}^{2}F^{G}
\end{equation*}

\begin{equation}
    g=\det (g_{GG^\prime{}})
\end{equation}
where $ G= \{ E,I \} $ represents excitatory $ E $ and inhibitory $ I $
processes, the aggregated relaxation time $ \tau $ is on the order of
10~ms, $ N=N^{E}+N^{I} $, and $ N^{E}=80 $, $ N^{I}=30 $ has been used
the number of $ E $ and $ I $ neurons in a minicolumn, with twice these
numbers for visual cortex. $ V^\prime{} $ are derived mesocolumnar
nearest-neighbor (NN) interactions among minicolumns within
macrocolumns, which are consistent with observed times for diffusion of
localized information across columns, and consistent with common time
scales of communication via diffusion of non-myelinated short-ranged
columnar fibers and via direct firings of myelinated long fibers across
regions, $ g^{G} $ are the drifts. $ g^{GG^\prime{}} $ is the covariance
matrix, the inverse of the metric $ g_{GG^\prime{}} $.

The threshold factor $ F^{G} $ is derived as

\begin{equation*}
    F^{G}=\sum \limits_{G^\prime{}}\frac{\nu ^{G}+\nu ^{\ddagger E^\prime
    {}}}{\big((\pi /2)[(v_{G^\prime{}}^{G})^{2}+(\phi _{G^\prime{}}^{G})^
    {2}](\delta ^{G}+\delta ^{\ddagger E^\prime{}})\big)^{1/2}}
\end{equation*}

\begin{equation*}
    \nu ^{G}=V^{G}-a_{G^\prime{}}^{G}v_{G^\prime{}}^{G}N^{G^\prime{}}-\frac
    {1}{2}A_{G^\prime{}}^{G}v_{G^\prime{}}^{G}M^{G^\prime{}}
\end{equation*}

\begin{equation*}
    \nu ^{\ddagger E^\prime{}}=-a_{E^\prime{}}^{\ddagger E}v_{E^\prime{}}^
    {E}N^{\ddagger E^\prime{}}-\frac{1}{2}A_{E^\prime{}}^{\ddagger E}v_{E^\prime
    {}}^{E}M^{\ddagger E^\prime{}}
\end{equation*}

\begin{equation*}
    \delta ^{G}=a_{G^\prime{}}^{G}N^{G^\prime{}}+\frac{1}{2}A_{G^\prime{}}^
    {G}M^{G^\prime{}}
\end{equation*}

\begin{equation*}
    \delta ^{\ddagger E^\prime{}}=a_{E^\prime{}}^{\ddagger E}N^{\ddagger
    E^\prime{}}+\frac{1}{2}A_{E^\prime{}}^{\ddagger E}M^{\ddagger E^\prime
    {}}
\end{equation*}

\begin{equation}
    a_{G^\prime{}}^{G}=\frac{1}{2}A_{G^\prime{}}^{G}+B_{G^\prime{}}^{G}\:,\:a_
    {E^\prime{}}^{\ddagger E}=\frac{1}{2}A_{E^\prime{}}^{\ddagger E}+B_{E^\prime
    {}}^{\ddagger E}
\end{equation}
where $ A_{G^\prime{}}^{G} $ and $ B_{G^\prime{}}^{G} $ are
minicolumnar-averaged inter-neuronal synaptic efficacies, $ v_{G^\prime{}}^
{G} $ and $ \phi _{G^\prime{}}^{G} $ are averaged means and variances of
contributions to neuronal electric polarizations. $ M^{G^\prime{}} $ and
$ N^{G^\prime{}} $ in $ F^{G} $ are afferent macrocolumnar firings,
scaled to efferent minicolumnar firings by $ N/N*\approx 10^{-3} $,
where $ N* $ is the number of neurons in a macrocolumn, about $ 10^{5} $.
Similarly, $ A_{G}^{G^\prime{}} $ and $ B_{G}^{G^\prime{}} $ have been
scaled by $ N*/N\approx 10^{3} $ to keep $ F^{G} $ invariant. Other
values taken are consistent with experimental data, e.g., $ V^{G}=10 $~mV,
$ v_{G^\prime{}}^{G}=0.1 $~mV, $ \phi _{G^\prime{}}^{G}=0.03^{1/2} $~mV.

The numerator of $ F^{G} $ contains post-synaptic parameters, and that
the denominator of $ F^{G} $ contains pre-synaptic parameters, a result
that drops out of the derivation of the mesoscopic derivation from the
statistics of synaptic and neuronal interactions in and across
minicolumns. Afferent contributions from $ N^{\ddagger E} $ long-ranged
excitatory fibers, e.g., cortico-cortical neurons, are included (implicitly
having coefficients measuring the strength of coupling between regions),
where $ N^{\ddagger E} $ might be on the order of 10\% of $ N^{*} $: Of
the approximately $ 10^{10} $ to $ 10^{11} $ neocortical neurons,
estimates of the number of pyramidal cells range from 2/3 up to 4/5
\citep{Markram+Toledo-Rodriguez+Wang+Gupta+Silberberg+Wu2004}. Nearly
every pyramidal cell has an axon branch that makes a cortico-cortical
connection, i.e., the number of cortico-cortical fibers is of the order $
10^{10} $. This development is used in the SMNI description of scalp EEG
across regions.
\subsubsection{Euler-Lagrange equations}
The Lagrangian components and Euler-Lagrange (EL) equations are
essentially the counterpart to classical dynamics,

\begin{equation*}
    \mathrm{Mass}=g_{GG^\prime{}}=\frac{\partial ^{2}L}{\partial (\partial
    M^{G}/\partial t)\partial (\partial M^{G^\prime{}}/\partial t)}
\end{equation*}

\begin{equation*}
    \mathrm{Momentum}=\Pi ^{G}=\frac{\partial L}{\partial (\partial M^{G}/\partial
    t)}
\end{equation*}

\begin{equation*}
    \mathrm{Force}=\frac{\partial L}{\partial M^{G}}
\end{equation*}

\begin{equation}
    \mathrm{F-ma}=0:\:\delta L=0=\frac{\partial L}{\partial M^{G}}-\frac
    {\partial }{\partial t}\frac{\partial L}{\partial (\partial M^{G}/\partial
    t)}
\end{equation}
Concepts like momentum, force, inertia, etc., are so ingrained into our
culture, that we apply them to many stochastic systems, like weather,
financial markets, etc., often without giving much thought to how these
concepts might be precisely identified. For a large class of stochastic
systems, even including nonlinear nonequilibrium multivariate
Gaussian-Markovian systems, like SMNI, the above formulation is precise.
For SMNI, this formulation has been particularly instructive
\citep{Ingber1983,Ingber+Nunez2010}. The Momentum defined above are also
used as Canonical Momentum Indicators (CMI) in several studies that
demonstrated its superiority over simple statistical correlations as
they take into account some physical properties of the systems studied.

The Fokker-Planck and path-integral representations present a compaction
of information relative to sources of noise. For example, consider the
set of Langevin stochastic differential equations in terms of Wiener
processes $ dW^{i} $, which can be rewritten in terms of m independent
sources of Gaussian noise $ \eta ^{i}=dW^{i}/dt $ (care is taken in the
limit).

\begin{equation}
    dM^{G}=f^{G}\big(t,M(t)\big)dt+\sum \limits_{i}\limits^{m}\hat{g}_{i}^
    {G}\big(t,M(t)\big)dW^{i}
\end{equation}
The mathematically equivalent Fokker-Planck and path-integral
representations
\citep{Ingber1991,Langouche+Roekaerts+Tirapegui1982}, develops a
covariance matrix where the $ m $ sources of noise are summed over,

\begin{equation}
    g^{GG^\prime{}}=\sum \limits_{i}\limits^{m}\hat{g}_{i}^{G}\hat{g}_{i}^
    {G^\prime{}}
\end{equation}
The CMI further compacts various sources of information. The CMI for a
single variable is the short-time deviation from the drift, divided by
the variance. The CMI for one of multiple variables is the short-time
deviation from the drift, multiplied by its components of the metric,
e.g., in the above context the $ G $ component $ g_{GG^\prime{}} $; if
the metric is diagonal, then this reduces to the inverse variance. The
CMI therefore contain information from both the drifts and diffusions.
For $ n $ variables, the symmetric covariance matrix may contain $ n(n+1)/2
$ components, which together with $ n $ drifts account for $ n(n+3)/2 $
sources of information. However, the CMI are $ n $ components, which is
useful for some systems.

The path-integral Lagrangian is not simply an alternative representation
of first and second moments similar to Langevin and Fokker-Planck
representations. In addition to possessing a variational principle, and
yielding indicators like canonical momenta, whereas the first and second
moments are generally accurate to order $ (\Delta t)^{1/2} $ (the
standard binomial tree algorithm is limited to Gaussian white noise
where the coefficient of the $ (\Delta t) $ term is zero), the
short-time propagator is a more accurate representation to order $ (\Delta
t)^{3/2} $
\citep{Langouche+Roekaerts+Tirapegui1982}, e.g., permitting a highly
robust binary tree path-integral algorithm to be developed for nonlinear
systems
\citep{Ingber+Chen+Mondescu+Muzzall+Renedo2001}.
\subsection{Coupling $ \mathrm{Ca}^{2+} $-waves with SMNI Lagrangian}
The SMNI approach is a bottom-up mesoscopic aggregation from microscopic
synaptic to columnar scales, and then scaled to relatively macroscopic
regional scales of neocortex, which has been further merged with larger
non-invasive EEG scales --- all at scales much coarser than molecular
scales. Here it is calculated how an SMNI vector potential (SMNI-VP)
constructed from magnetic fields induced by neuronal electrical firings,
at thresholds of collective minicolumnar activity with laminar
specification, can give rise to causal top-down mechanisms that affect
molecular excitatory and inhibitory processes in STM and LTM\@.

While many studies have examined the influences of changes in $ \mathrm{Ca}^
{2+} $ distributions on large-scale EEG
\citep{Kudela+Bergey+Franaszczuk2009}, future work will examine the
complimentary effects on $ \mathrm{Ca}^{2+} $ ions at a given neuron
site from EEG-induced magnetic fields arising from other neuron sites.
Here, sufficient calculations claim the importance of macroscopic EEG $
\mathbf{A} $, arising from microscopic synchronous neural activity, on
molecular momenta $ \mathbf{p} $ in $ \mathrm{Ca}^{2+} $ ions.

The time dependence of $ \mathrm{Ca}^{2+} $ wave momenta may be
calculated with rate-equations
\citep{Li+Rinzel1994} as a Hodgkin-Huxley model
\citep{Hodgkin+Huxley1952}, including contributions from astrocytes in
the vicinity of synapses
\citep{Bezzi+Gundersen+Galbete+Seifert+Steinhause+Pilati+Volterra2004,Larter+Craig2005}.
In this study, the resulting flow of $ \mathrm{Ca}^{2+} $ wave momenta
will be further determined by its interactions in $ \mathbf{\Pi } $, the
canonical momenta which includes $ \mathbf{A} $.

One influence of $ \mathrm{Ca}^{2+} $ likely is regulating synaptic
interactions
\citep{Manita+Miyazaki+Ross2011}. The SMNI Lagrangian explicitly
describes where the $ \mathrm{Ca}^{2+} $ affect the columnar-averaged
synaptic parameters $ \{ A_{G^\prime{}}^{G},B_{G^\prime{}}^{G},A_{E^\prime
{}}^{\ddagger E},B_{E^\prime{}}^{\ddagger E} \} $. In this context $
\mathrm{Ca}^{2+} $ wave activity can affect the $ A $ and $ B $ synaptic
parameters in these equations, while the $ \mathbf{A} $ EEG fields
affect the $ \mathrm{Ca}^{2+} $ waves.

In SMNI papers, the CM is invoked by fine-tuning $ B $ parameters to
bring maximum multiple minima in firing space $ M $, by adjusting
background $ B_{G^\prime{}}^{G} $ to set $ \nu ^{G}=0 $ when $ M^{G}=0 $,
similar to the control of spontaneous synaptic background observed
during selective attention
\citep{Briggs+Mangun+Usrey2013,Mountcastle+Andersen+Motter1981}. The
latter authors state that ``it is unclear how attention-mediated
alterations in neuronal population activity translate into improvements
in perception.'' In the SMNI context, the $ B $ parameters are a logical
first choice to include influences from columnar $ \mathrm{Ca}^{2+} $
activities. These minima tend to lie along a line in a trough in $ M $
space, essentially $ A_{E}^{E}M^{E}-A_{I}^{E}M^{I}\approx 0 $, noting
that in $ F^{I} $ $ I-I $ connectivity is experimentally observed to be
very small relative to other pairings, so that $ (A_{E}^{I}M^{E}-A_{I}^{I}M^
{I}) $ is typically small only for small $ M^{E} $. This model gives
rise to all the successful SMNI calculations describing various STM
phenomena.

This trough also supported previous SMNI work fitted to EEG data
\citep{Ingber1997}, developing a scaled macrocolumnar electric potential $
\Phi _{\nu } $ at scalp region $ \nu $ derived with first and second
moments of the SMNI Lagrangian, the argument of the associated
distribution $ P_{\nu } $, at each (interconnected) region $ \nu $,

\begin{equation*}
    P_{\nu }[\Phi _{\nu }(t)|\Phi _{\nu }(t-\Delta t)]=\frac{1}{(2\pi
    \sigma ^{2}\Delta t)^{1/2}}\exp (-L_{\nu }\Delta t)
\end{equation*}

\begin{equation*}
    L_{\nu }=\frac{1}{2\sigma ^{2}}(\dot{\Phi }_{\nu }-m)^{2}
\end{equation*}

\begin{equation*}
    m=<\Phi _{\nu }-<\phi >>=a<M^{E}>+b<M^{I}>=ag^{E}+bg^{I}
\end{equation*}

\begin{equation}
    \sigma ^{2}=<(\Phi _{\nu }-<\phi >)^{2}>-<\Phi _{\nu }-<\phi >>^{2}=a^
    {2}g^{EE}+b^{2}g^{II}
\end{equation}
in terms of $ M^{G} $-space drifts $ g^{G} $, diffusions $ g^{GG^\prime{}}
$, and an averaged reference $ <\phi > $.

The same process supports the similar parameterization of $ \mathbf{A} $
in these studies, i.e.,

\begin{equation}
    \mathbf{A}=cM^{E}\mathbf{\hat{r}}+dM^{E}\mathbf{\hat{r}}
\end{equation}
where $ c $ and $ d $ are scaled to the order of $ 10^{4} $~pA, as
discussed above. This results in a Lagrangian $ L $ for the combined
EEG-$ \mathrm{Ca}^{2+} $ system, e.g., considering $ |\mathbf{A}| $ as
primarily perpendicular to the scalp,

\begin{equation*}
    L=\frac{1}{2\sigma ^\prime{}^{2}}(|\dot{\mathbf{A}}|_{\nu }-m^\prime
    {})^{2}
\end{equation*}

\begin{equation*}
    m^\prime{}=<|\mathbf{A}|_{\nu }-<\phi ^\prime{}>>=c<M^{E}>+d<M^{I}>=cg^
    {E}+dg^{I}
\end{equation*}

\begin{equation}
    \sigma ^\prime{}^{2}=<(|\mathbf{A}|_{\nu }-<\phi ^\prime{}>)^{2}>-<|\mathbf
    {A}|_{\nu }-<\phi ^\prime{}>>^{2}=c^{2}g^{EE}+d^{2}g^{II}
\end{equation}
where now $ m^\prime{} $ and $ \sigma ^\prime{} $ are nonlinear
functions of $ \mathbf{A} $ via the $ B $ synaptic parameters discussed
further below. This Lagrangian is the argument of the exponential
defining the conditional probability density for developing from a state
at time $ t-1 $ to time $ t $. The variational principle obeyed by this
Lagrangian permits optimization of parameters to find most likely states
that best fit EEG data, i.e., including macrocolumnar parameters within
regions, long-ranged connectivity and time delays across regions
\citep{Ingber1997}.
\subsection{Experimental verification}
The duration of a $ \mathrm{Ca}^{2+} $ wave can be on the order of
500~ms, so that the momenta of such ions can be importantly influenced
during relatively long EEG events like N100 and P300 potentials,
reflecting latencies on the order of 100~ms and 300~ms, common in
selective attention tasks which span these events
\citep{Srinivasan+Winter+Ding+Nunez2007}. Similar to procedures used in
previous SMNI fits to EEG data
\citep{Ingber1997,Ingber1998}, here the influence of $ \mathrm{Ca}^{2+} $
waves may be tested by parameterizing the $ B $ synaptic parameters to
include their influence in data sets where subjects have had
simultaneous recording of scalp EEG and samplings of $ \mathrm{Ca}^{2+} $
wave activity at synaptic scales. These parameters are then fit to a
portion of the EEG data, the in-sample set. The trained parameters can
be used in out of sample EEG data, to test if the included $ \mathrm{Ca}^
{2+} $ wave activity correlates with the observed $ \mathrm{Ca}^{2+} $
wave data.

The interaction of $ \mathbf{A} $ and $ \mathrm{Ca}^{2+} $ waves can be
detailed using SMNI-scaled synaptic parameters which include a term
dependent on $ \mathbf{A} $, with coefficients measuring the convergence
of synaptic interactions from many local minicolumnar and regional
long-ranged fibers. The waves depend on aggregates of their $ \mathbf{\Pi
}=\mathbf{p}+q\mathbf{A} $ dynamics. E.g., this can be modeled as a
Taylor expansion in $ |\mathbf{A}| $,

\begin{equation}
    B_{G^\prime{}}^{G}\rightarrow B_{G^\prime{}}^{G}+|\mathbf{A}|B^\prime
    {}_{G^\prime{}}^{G}\:,\:B_{E^\prime{}}^{\ddagger E}=B_{E^\prime{}}^{\ddagger
    E}+|\mathbf{A}|B^\prime{}_{E^\prime{}}^{\ddagger E}
\end{equation}
Eventually, the functional form of these dynamics should be established
by models fit to molecular dynamics data, but for now at least their
parameterized influences can be included. Since $ \Phi $ is
experimentally measured, not $ \mathbf{A} $, but both are developed by
currents $ \mathbf{I} $, when fitting to EEG data, it is reasonable to
consider $ \mathbf{A} $ as proportional to $ \Phi $ with a simple
scaling factor, and now the additional parameterization of $ B_{G^\prime
{}}^{G} $ and $ B_{E^\prime{}}^{\ddagger E} $ are to be included to
modify previous work. To handle the otherwise recursive calculation of $
|\mathbf{A}| $ multiplying $ B^\prime{}_{G^\prime{}}^{G} $ and $ B^\prime
{}_{E^\prime{}}^{\ddagger E} $, here $ |\mathbf{A}| $ is saved as a
multiple of $ |g^{GG}|\tau $ from just-previous data points, to be used
in current time in the cost function calculation. A reasonable
constraint is imposed that the inclusion of the $ B^\prime{} $ terms not
exceed the value of the $ B $ terms, e.g., limiting the influence of $ B^\prime
{} $ to at most doubling the background noise. The data used for this
study, is spaced about 3.6~ms ($ <\tau $) between 150-400~ms after
presentation of stimuli
\citep{Ingber1997,Ingber1998}.

The values of averaged synaptic parameters used in the 1980's SMNI
papers were taken from experimental papers. Without any fitting of these
parameters to other data, SMNI detailed STM phenomena, e.g., as
mentioned previously, capacity (auditory $ 7\pm 2 $ and visual $ 4\pm 2 $),
duration, stability, primacy versus recency rule, observed times for
diffusion of localized information across columns, Hick's law --- and
other properties of neocortex by scaling up to macrocolumns across
regions to fit EEG data. The $ B $'s terms, previous to the present $ B^\prime
{} $ inclusion, were consistent with these STM observations. Therefore,
in this study, the $ B^\prime{} $ terms were constrained to add no more
than their $ B $ counterparts. Furthermore, since the data being fit is
within the duration of P300 EEG waves, the inclusion of the
time-dependent $ B^\prime{} $ terms, i.e., including time-dependent
modeled $ |\mathbf{A}| $, required a ``dynamic centering mechanism'' (DCM)
to model regular access to maximum memory states, consistent with the
early SMNI studies. Other future studies mentioned below will simulate
the contribution of $ \mathrm{Ca}^{2+} $ waves via tripartite synaptic
interactions, to determine if the changes implemented are reasonable
assumptions.

Another experimental test at the classical molecular scale to verify the
influence of $ \mathbf{A} $, can be made considering that if the current
lies along $ \mathbf{\hat{z}} $, then $ \mathbf{A} $ only has components
along $ \mathbf{\hat{z}} $, and

\begin{equation}
    \mathbf{\Pi }=p_{x}\mathbf{\hat{x}}+p_{y}\mathbf{\hat{y}}+(p_{z}+qA_
    {z})\mathbf{\hat{z}}
\end{equation}
The influence of time-dependent $ \mathrm{Ca}^{2+} $ waves is introduced
in the post-synaptic and pre-synaptic SMNI parameters, which here also
are time-dependent as functions of changing $ \mathrm{Ca}^{2+} $ ions.
Such parameters are present at neuronal scales and are included in
microscopic ordinary differential equation calculations. However, as in
the original development of SMNI, these parameters are developed to
mescolumnar scales, and the prediction here is that there is a
predominance of $ \mathrm{Ca}^{2+} $ waves in directions closely aligned
to the direction perpendicular to neocortical laminae during strong
collective EEG\@.

The prediction above is made within the context of classical physics,
due to the large value of $ \mathbf{A} $ relative to $ \mathbf{p} $
during periods of large synchronous columnar firings. Note that the
context of the quantum physics calculations above also is relevant, even
within short coherence times, since the bias of $ \mathbf{A} $ is
present at the very earliest of times when $ \mathbf{p} $ is
appreciable.
\section{Current projects}
Several sub-projects are being developed, using codes that permit some
parallelization.
\subsection{EEG data fits}
EEG data is fit to SMNI, using data from \hfil\\
http://kdd.ics.uci.edu/databases/eeg/ that the author has made public, and
which is regularly used by other researchers. This project examines the
influence of $ \mathbf{A} $ on the $ B $ synaptic parameters in the SMNI
Lagrangian.

\subsubsection{Data}
EEG spontaneous and evoked potential (EP) data from a multi-electrode
array under a variety of conditions was collected at several centers in
the United States, sponsored by the National Institute on Alcohol Abuse
and Alcoholism (NIAAA) project
\citep{Zhang+Begleiter+Porjesz1997,Zhang+Begleiter+Porjesz+Litke1997,Zhang+Begleiter+Porjesz+Wang+Litke1995}.
This data set was used in earlier SMNI studies
\citep{Ingber1997,Ingber1998}. These experiments, performed on carefully
selected sets of subjects, suggest a genetic predisposition to
alcoholism that is strongly correlated to EEG AEP responses to patterned
targets.

The ASA code is used for fitting SMNI to this data
\citep{Ingber1993}. These fits permit an estimate of the influence of $ |\mathbf
{A}| $ on the $ B $ synaptic parameters.

It is clear that the authors are not an expert in the clinical aspects
of these alcoholism studies. It suffices for this study that the data
used is clean raw EEG data, and that these SMNI, CMI, and ASA techniques
can and should be used and tested on other sources of EEG data as well.

Each set of results is presented with 6 figures, labeled as [\{alcoholic
$ | $ control\}, \{stimulus 1 $ | $ match $ | $ no-match\}, subject, \{potential
$ | $ momenta\}], abbreviated to \{a $ | $ c\}\_\{1 $ | $ m $ | $ n\}\_subject
where match or no-match was performed for stimulus 2 after 3.2 sec of a
presentation of stimulus 1
\citep{Zhang+Begleiter+Porjesz1997,Zhang+Begleiter+Porjesz+Litke1997,Zhang+Begleiter+Porjesz+Wang+Litke1995}.
Data includes 10 trials of 69 epochs each between 150 and 400 msec after
presentation. For each subject run, after fitting 28 parameters with
ASA, using the training data, epoch by epoch averages are developed of
the raw data and of the multivariate SMNI CMI. It was noted that much
poorer fits were achieved when the CM
\citep{Ingber1984,Ingber1985}, driving multiple attractors into the
physical firing regions bounded by $ M^{G}\le \pm N^{G} $, was turned
off and the denominators in $ F^{G} $ were set to constants, confirming
the importance of using the full SMNI model. All stimuli were presented
for 300 msec. Note that the subject number also includes the \{alcoholic
$ | $ control\} tag, but this tag was added just to aid sorting of files
(as there are contribution from co2 and co3 subjects). Each figure
contains graphs superimposed for 6 electrode sites (out of 64 in the
data) which have been modeled by SMNI using a circuitry of frontal sites
(F3 and F4) feeding same-side sites: temporal sites (T7 and T8) with
delay times of 1 unit of data resolution, parietal sites (P7 and P8)
with delay times of 2 units of data resolution. Temporal sites also feed
same-side parietal sites with delay times of 1 unit of data resolution.
Additionally, there are cross-side interactions, between temporal sites
and between parietal sites with delay times of 1 unit of data
resolution. Odd-numbered (even-numbered) sites refer to the left (right)
brain.

Parameters to be fit to EEG are in the EEG Lagrangian $ L_{\nu } $ at
each regional electrode site $ \nu $, explicitly as \{$ <\phi > $, $ a $,
$ b $\}, and through the SMNI threshold factor $ F^{G} $ implicitly as
\{$ B1 $, $ d_{\nu } $\}, where $ B1 $ is parameterized in the $ \mathbf
{A} $ model but set to a constant for the no-$ \mathbf{A} $ model, and
the $ d_{\nu } $ are coefficients representing strengths of coupling
across regional sites to a given region.
\subsubsection{Optimization}
This optimization used ASA, 2013 version 28.15, tuned to give reasonable
performance by examining intermediate results of several sample runs in
detail. See the ASA code for a discussion of ASA OPTIONS and Tuning
\citep{Ingber1993,Ingber2012c}.

For both $ \mathbf{A} $ and no-$ \mathbf{A} $ models, with DCM instead
of just CM in previous studies, ASA was used for 60 data sets in \{a\_n,
a\_m, a\_n, c\_1, c\_m, c\_n\} of 10 subjects. Each of these data sets
had 4-6 parameters for each SMNI electrode-site model in \{F3, F4, T7,
T8, P7, P8\}, i.e., 34 parameters (28 parameters when $ B1 $ terms are
dropped) for each of the optimization runs, to be fit to over 400 pieces
of potential data. This again is the same procedure used in previous
papers with this data
\citep{Ingber1997,Ingber1998}.

The ranges of the parameters were decided as follows. The ranges of the
strength of the long-range connectivities $ d_{\nu } $ were from 0 to 1.
The ranges of the $ \{ a,b,\phi \} $ parameters were decided by using
minimum and maximum values of $ M^{G} $ and $ M^{\ddagger G} $ firings
to keep the potential variable within the minimum and maximum values of
the experimentally measured potential at each electrode site.

It was found that typically within several thousand generated states,
the global minimum was approached within at least one or two significant
figures of the effective Lagrangian (including the prefactor). This
estimate was based on fits achieved with 2,000,000 generated states per
run, after which ASA used its supplementary simplex code for additional
local fits to sometimes get tighter fits. Each run generates 10 files,
of type asa-out, asa-usr-out, train-run-out, test-run-out, train-data,
test-data, train-test-graph.ps, train-test-graph.eps,
train-test-graph.pdf, and train-test-graph.png. Each ASA optimization
took about 6 CPU-hrs for each of 120 runs on the XSEDE (www.xsede.org)
Trestles cluster, a cumulative CPU-month+ for 1200 files. An additional
set of runs used 4,000,000 generated states per run, which improve a few
graphs; this set is given here.

Each complete set of runs was performed on the XSEDE Trestles machine,
using four nodes, each node spawning 30 processors. Trestles is a
dedicated XSEDE cluster designed by Appro and SDSC consisting of 324
compute nodes, running under Linux CentOS.  Each compute node contains
four sockets, each with a 8-core 2.4 GHz AMD Magny-Cours processor, for
a total of 32 cores per node and 10,368 total cores for the system.
Each node has 64 GB of DDR3 RAM, with a theoretical memory bandwidth of
171 GB/s.  The compute nodes are connected via QDR InfiniBand
interconnect, fat tree topology, with each link capable of 8 GB/s (bidirectional).
Trestles has a theoretical peak performance of 100 TFlop/s.
\subsubsection{Testing data}
When the parameters of a theory of a physical system possess clear
relationships to observed physical entities, and the theory fits
experimental phenomenon while the parameters stay within experimentally
determined ranges of these entities, then generally it is conceded that
the theory and its parameters have passed a reasonable test. It is
argued that this is the case for SMNI and its parameters, and this
approach sufficed for the first study of the present data
\citep{Ingber1997}, just as SMNI also has been tested in previous papers.

When a model of a physical system has a relatively phenomenological
nature then often such a model is best tested by first ``training'' its
parameters on one set of data, then seeing to what degree the same
parameters can be used to match the model to out-of-sample ``testing''
data. For example, this is performed for the statistical mechanics of
financial markets (SMFM) project, applied to trading models, as
documented on http://www.ingber.com .

In the present project, there exists barely enough data to additionally
test SMNI in this training versus testing methodology. That is, when
first examining the data, it was decided to to try to find sets of data
from at least 10 control and 10 alcoholic subjects, each set containing
10 runs for each of the 3 experimental paradigms, as reported in a
previous paper
\citep{Ingber1997}. When reviewing this data, e.g., for the example of
the one alcoholic and the one control subject which were illustrated in
graphs in that previous paper, it was determined that there exists 10
additional sets of data for each subject for each paradigm, except for
the c\_n case of the no-match paradigm for the control subject where
only 5 additional out-of-sample runs exist. For this latter case, to
keep the number of runs sampled consistent across all sets of data,
e.g., to keep the relative amplitudes of fluctuations reasonably
meaningful, 5 runs of the previous testing set were joined with the 5
runs of the present training set to fill out the data sets required for
this study.
\subsubsection{Graphical results}
The above procedure was followed both when including $ \mathbf{A} $
terms in the synaptic background parameters in the SMNI model, and when
excluding these terms (the previously published SMNI model).

The top graphs in each set are the EEG potentials; each channel line in
the graphs represents an average over 10 trials. The bottom graphs in
each set are the CMI calculated from the fitted Lagrangian. The left
graphs in each set are the training sets. The right graphs in each set
are the test sets.

With the $ \mathbf{A} $ terms, for both Training and Testing, Fig.  1
compares the CMI to raw data for an alcoholic subject for the a\_m
paradigm, for both the training and testing data. The CMI have been
truncated to lie with [-0.5, 0.5] to keep the scale exposing their
structure; an occasional small volatility can cause a point to jump out
of this scale. To facilitate comparisons, the representative subjects
presented here were selected to be the same as those presented in
original studies
\citep{Ingber1997,Ingber1998}. In previous papers, the scale was
permitted to vary for each graph, which in some cases served well to
detail some trends, but in other cases made relative comparisons
difficult. Also, advantage has been taken of powerful computer resources
in this project, to greatly increase the optimization of parameters
relative to previous studies, and permitting more CPU-intensive adaptive
DCM at each epoch in all runs, whereas only CM was used at the first
epoch in previous studies.

Note that the CMI are essentially short-time deviations from drifts
divided by variances, and therefore may have different ``shapes'' than
the raw data; the CMI do not track the EEG data. In this context, it is
becoming more common to develop indicators of signals in multi-channel
data that do not necessarily track the raw data
\citep{Roy+Schaffer+Laramee2013}. The SMNI CMI generally have tighter
``shapes'' than the raw data signifying tighter synchronization among
the recorded channels; previous papers gave more numerical details. The
CMI give better signal to noise resolution than the raw data, especially
comparing the significant matching tasks between the control and the
alcoholic groups, e.g., the c\_m and a\_m paradigms, in both the
training and testing cases. The $ \mathbf{A} $ model barely shows
tighter synchronization among the channels and more pronounced signal
near the end of the epoch than does the no-$ \mathbf{A} $ model. While
these features are barely visible via the CMI for individual subjects,
they are quite visibly apparent in the aggregated data in Fig.  2 where
data from all subjects and paradigms, representing 11,075 sessions, are
aggregated in histograms developed by the UCI Knowledge Discovery in
Databases (KDD) staff. (KDD is now merged with the UCI Machine Learning
Repository at http://archive.ics.uci.edu/ml/ .) \hfil\\
See http://kdd.ics.uci.edu/databases/eeg/alcoholic.gif and \hfil\\
http://kdd.ics.uci.edu/databases/eeg/control.gif for this comparison, as
described in the \hfil\\
http://kdd.ics.uci.edu/databases/eeg/eeg.data.html introduction to the EEG
database.

While Laplacian filtering of EEG data can help determine localization of
activity under electrode sites
\citep{Nunez+Srinivasan2006}, fitting the circuitry in the SMNI
Lagrangian also helps to serve this purpose, thereby enhancing the
information that modeling can extract from the raw data
\citep{Ingber1997,Ingber1998}.

Tables I and II provide \{c\_1, c\_n, c\_m, a\_1, a\_n, a\_m\} runs
versus all fitted parameters, 34 for the $ \mathbf{A} $ model and 28 for
the no-$ \mathbf{A} $ model, resp., over electrode sites \{F3, F4, T7,
T8, P7, P8\}. These tables are simply means over 10 subjects per each
paradigm, and accordingly have wide variances. Nevertheless, they serve
to identify the magnitudes of the parameters that are optimized per
subject.

\pagebreak[4]
\begin{center}
    \includegraphics[width=6in]{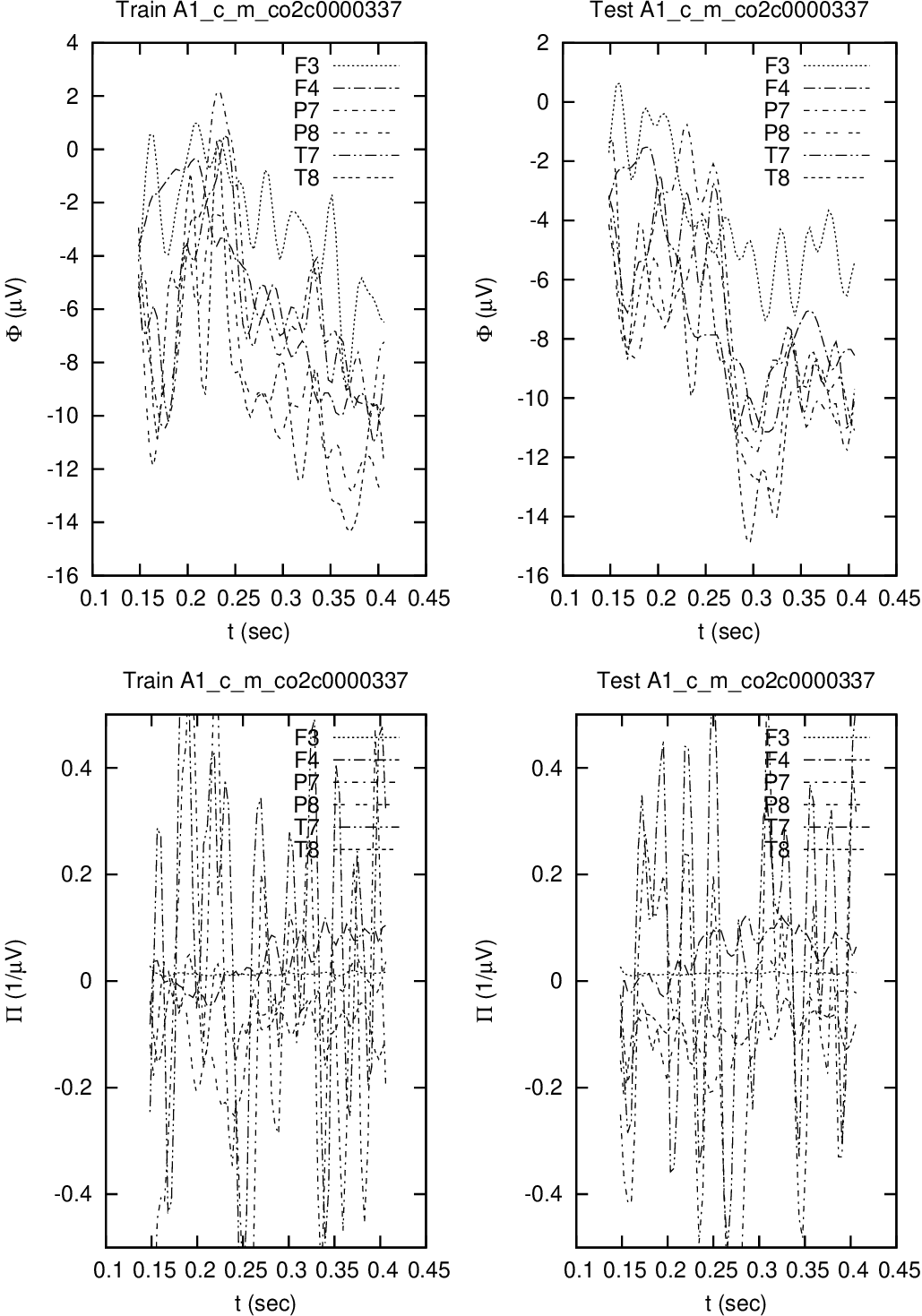}
\end{center}

\begin{quote}
    FIG.  1. With the $ \mathbf{A} $ model, the match second-stimulus c\_m
    paradigm for control subject co2c0000337, plots are given of
    activities under 6 electrodes of the CMI in the lower figures, and
    of the electric potential in the upper figures.
\end{quote}

\pagebreak[4]
\begin{center}
    \includegraphics[width=6in]{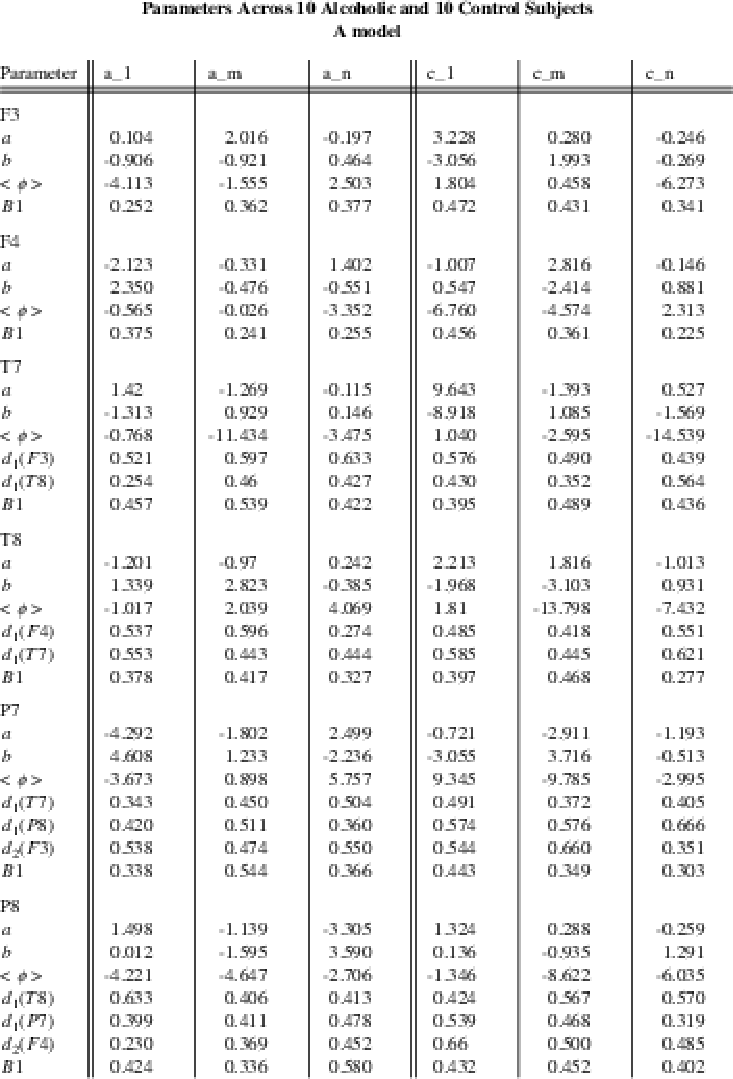}
\end{center}

\begin{quote}
    Table.  I. Parameters across 10 alcoholic and 10 control subjects
    are given for the $ \mathbf{A} $ model.
\end{quote}

\pagebreak[4]
\begin{center}
    \includegraphics[width=6in]{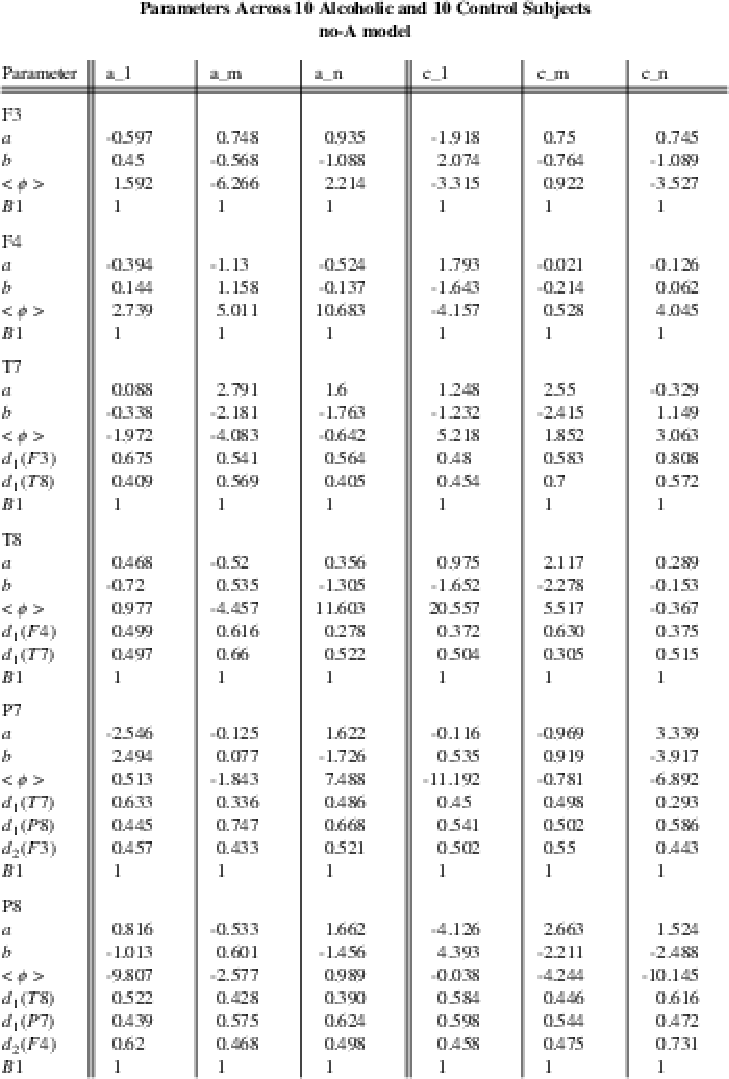}
\end{center}

\begin{quote}
    Table.  II. Parameters across 10 alcoholic and 10 control subjects
    are given for the no-$ \mathbf{A} $ model.
\end{quote}

Similar results are seen for other 10 control and 10 alcoholic subjects.
For each set of $ |\mathbf{A}| $ and no-$ |\mathbf{A}| $ runs, there are
360 files of output data, 240 files of 4 graphs each (in ps, eps, png,
and pdf formats), 80 files of parameter tables, and 13 files of summary
statistics.

After the above training-testing methodology is applied to more
subjects, with additional variants of $ \mathbf{A} $ models, it will
then be possible to perform additional statistical analyses to seek more
abbreviated measures of differences between alcoholic and control groups
across the 3 experimental paradigms.
\subsubsection{Supplementary analysis}
The first author (LI) is responsible for this main text, the computer
codes, and graphs of EEG data and CMI data as presented here. The other
authors (RS and MP) are responsible for their Supplementary analysis
included in the Supplementary material, which also contains all the
graphs.

The amount of EEG data per subject per paradigm is slim. Also, it is
relevant in the context of this project to investigate potential
clinical use of this work. Therefore, in addition to usual analyses,
e.g., including moments of fits, etc., ``perceived'' results that are
reasonably visually obvious are included. This required at least as much
work as usual analyses.

Overall, mixed results were demonstrated when attempting to evaluate the
efficacy or improvements of the CMI when compared to the EEG data.
However, many definitively positive improvements with the $ \mathbf{A} $
model were observed, both when comparing to the EEG data and the no-$
\mathbf{A} $ model. Three levels of analysis were performed.

(A) Presentation of the supplemental analysis of the data, sorted by
subject, with all six channels displayed at once, was organized as
previously detailed and in the original study. The $ \mathbf{A} $ model
performed best in 51.6\% of cases, worse in 28.3\%, with the remaining
20\% inconclusive. Within the worse data set, the no-$ \mathbf{A} $
model performed better than the $ \mathbf{A} $ model in two cases.
Significant improvements in separation of signals, moderate improvements
in synchrony, and a general reduction in amplitude and frequency of the
CMI are readily observable and present in the majority of all paradigms
\{1 $ | $ m $ | $ n\} and across both groups \{alcoholic $ | $ control\}
with the $ \mathbf{A} $ model in comparison to performing the CMI
calculations with the no-$ \mathbf{A} $ model.

The $ \mathbf{A} $ model produces near to visibly totally flat waves in
over half of all subjects, with no obvious link to a specific group or
paradigm. These flat waves are almost always only present in one wave
per case, at or very near the y-origin, with a very few cases showing
more than one flat or near-flat wave. However, the $ \mathbf{A} $ model
provided clearer insight and cleaner representation of the EEG data than
the EEG plots themselves when presented in this manner.

(B) When comparing paradigms on each plot together, the $ \mathbf{A} $
model shows significant improvement over the no-$ \mathbf{A} $ model in
most cases, exhibiting general increase of separation of signals, as
well as preserving the relative positive shift of the n paradigm above
the m paradigm. The $ \mathbf{A} $ model outperforms the EEG plots when
discerning between m and n paradigms for channels F4, P7 and P8 within
the test data set across both groups, as well as increases the ability
to discern alcoholic versus control subjects when comparing paradigms
for the same channels.

When the data are sorted by individual paradigm and only the two CMI
models compared, the $ \mathbf{A} $ model shows the best performance
within paradigm m compared to the no-$ \mathbf{A} $ model. With the
remaining 1 and n paradigms, the results are mixed and ambiguous,
showing very little improvement with the $ \mathbf{A} $ model. Markedly
worse performance in channel T8 across both paradigms \{1 $ | $n\} and
additionally worse performance in channel T7 with the 1 paradigm are
evident with the $ \mathbf{A} $ model. The $ \mathbf{A} $ model performs
better than the no-$ \mathbf{A} $ model when specifically discerning
alcoholic from control groups within the n paradigm.

When comparing the three paradigms, including the EEG data in the
analysis, the $ \mathbf{A} $ model performs best over the no-$ \mathbf{A}
$ model and EEG plots in one case, specifically the Test data for the
control group, among channels F4, P7 and P8.  In the remaining cases,
the EEG data showed best.

Figs.  3-5 give perceived relative improvements among EEG data, $
\mathbf{A} $ and no-$ \mathbf{A} $ models, for various paradigms.

(C) The final analysis on the data was performed with the channels
recombined resulting in three-dimensional plots.  Overall, there were
expected differences between Train and Test data within the CMI plots.

When attempting to discern differences between alcoholic and control
data, the $ \mathbf{A} $ model performed best within the m paradigm,
Train and Test data sets.  The remaining cases were handled best by the
EEG plots, with one exception of the Test data for paradigm 1 as being
inconclusive.  Within these cases, the $ \mathbf{A} $ model typically
performed better than no-$ \mathbf{A} $, with one notable exception
being the train, control data set for the n paradigm.

When examining these last plots comparing the three paradigms, the EEG
plots performed best in almost all cases with one notable exception.
The $ \mathbf{A} $ model performed best distinguishing the m and n
paradigms within the train data for the alcoholic group.

\pagebreak[4]
\begin{center}
    \includegraphics[width=6in]{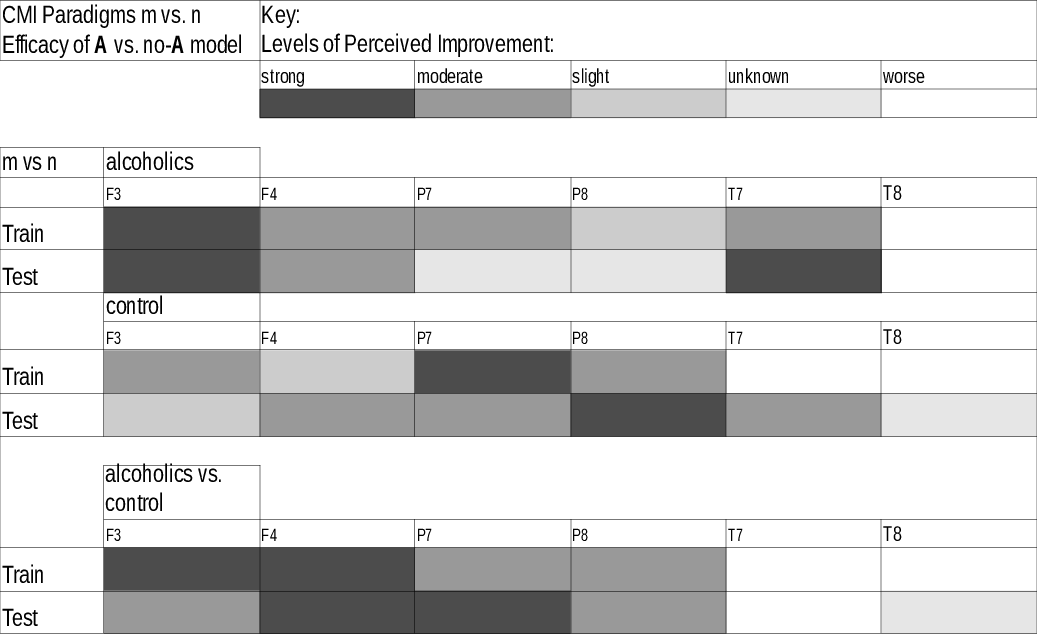}
\end{center}

\begin{quote}
    FIG.  3. Chart of perceived relative improvement of CMI for
    paradigms m and n, $ \mathbf{A} $ versus no-$ \mathbf{A} $ models
\end{quote}

\pagebreak[4]
\begin{center}
    \includegraphics[width=6in]{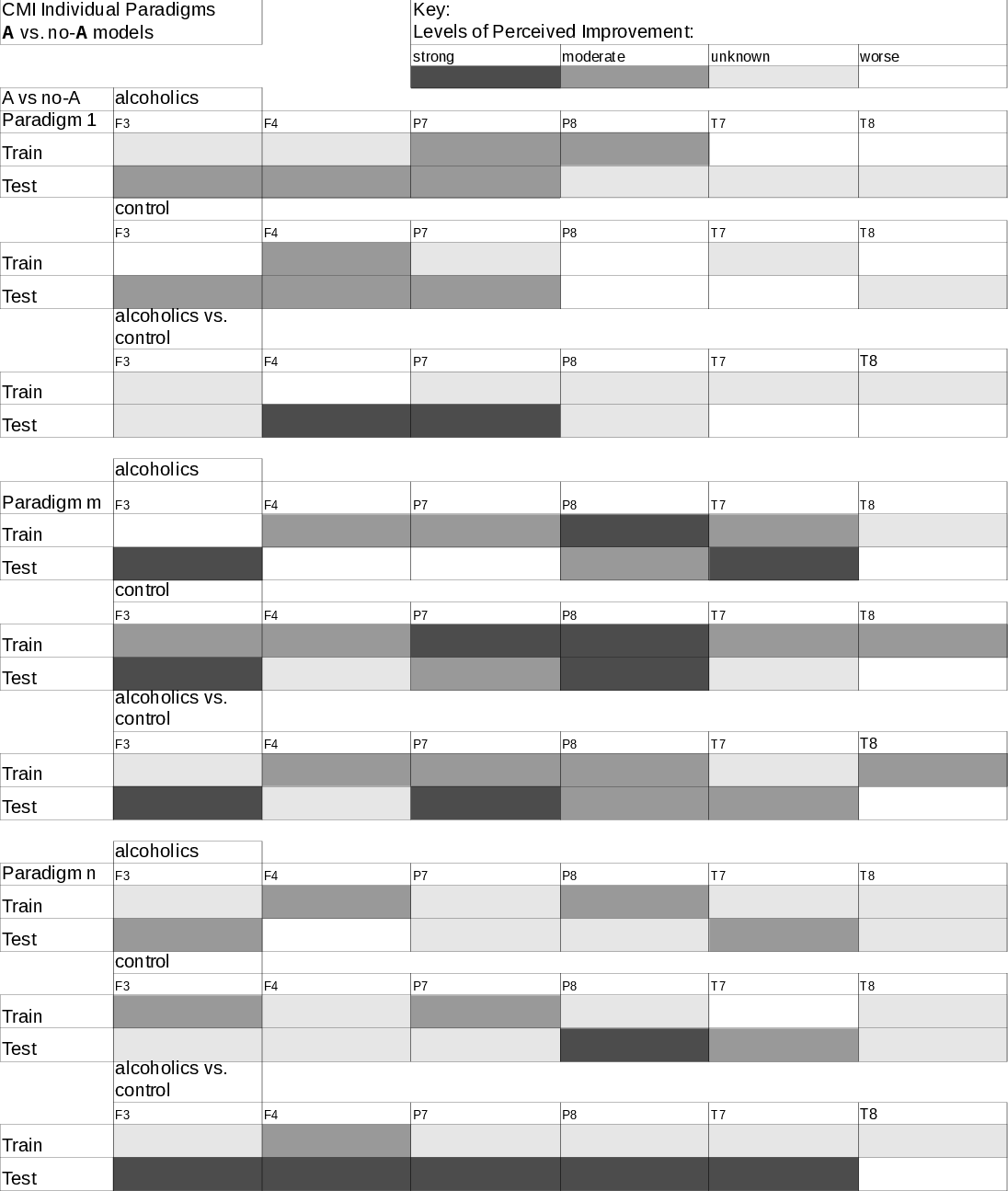}
\end{center}

\begin{quote}
    FIG.  4. Chart of perceived relative improvement of Individual CMI
    paradigms, $ \mathbf{A} $ versus no-$ \mathbf{A} $ models
\end{quote}

\pagebreak[4]
\begin{center}
    \includegraphics[width=6in]{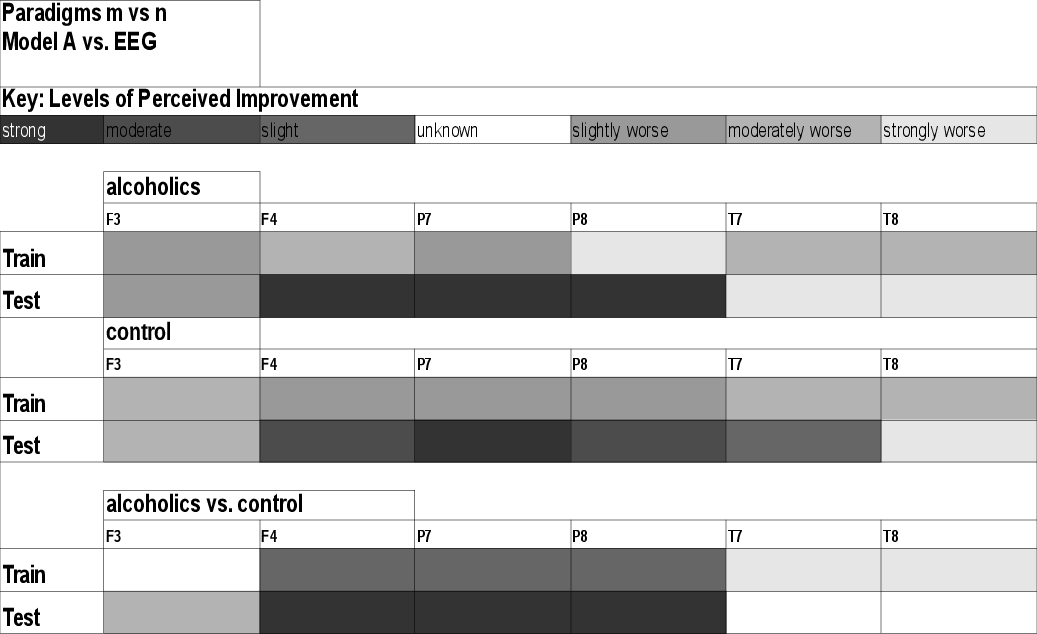}
\end{center}

\begin{quote}
    FIG.  5. Chart of perceived relative improvement of $ \mathbf{A} $
    model versus EEG, for paradigms m and n
\end{quote}
\clearpage
\subsubsection{Discussion}
While the CMI show marked differences between the $ \mathbf{A} $ and no-$
\mathbf{A} $ runs, as discussed above, the degrees of fitting, measured
by the values of the cost functions $ L $, are close. No strong
conclusion can be drawn for these runs regarding the superiority of
either model. This likely is due to the constraint imposed on the $
\mathbf{A} $ model, that the inclusion of the $ B^\prime{} $ terms not
exceed the value of the $ B $ terms, e.g., limiting the influence of $ B^\prime
{} $ to at most doubling the background noise; this saturation happened
quite often. Other $ \mathbf{A} $ models are being tested.

This SMNI model and associated algorithms serve to test the influence of
molecular $ \mathrm{Ca}^{2+} $-wave processes at regional scales at the
level of scalp EEG\@. Such models are important to test multiple-scale
processes in neocortex.
\subsection{Contribution of $ \mathrm{Ca}^{2+} $ waves}
We would develop a simple model that considers $ \mathrm{Ca}^{2+} $
waves that trigger glutamate release, as an estimate of the importance
of tripartite $ \mathrm{Ca}^{2+} $ waves to the the $ B $ synaptic
parameters as derived in SMNI papers
\citep{Ingber1982,Ingber1983}. In those 1982 and 1983 papers, derivations
based on different distributions for the number of quanta $ q $ released
across synaptic gaps, gave the same means for the distribution of $ q $,
represented by $ a_{G^\prime{}}^{G} $ and $ a_{E^\prime{}}^{\ddagger E} $.
The $ a $'s are a sum of $ 1/2A $'s due to neuronal firings plus
background $ B $'s due to background noise. It has been observed that
changes in [$ \mathrm{Ca}^{2+} $] appear to influence release of
glutamate and postsynaptic firing
\citep{Sharma+Vijayaraghavan2003}. It is reasonable to consider that $
\mathrm{Ca}^{2+} $ waves from tripartite interactions contribute to the $
B $'s.

There are several published models and codes that relate to our project,
but not quite closely enough to simply use in their present form.
\subsubsection{Generic simulators}
There are generic brain simulators that make is easier to patch together
stochastic equations and with reasonable experimentally determined
parameters to construct new models.

The NEURON code is a well-known simulator available on our xsede.org
platforms
\citep{Carnevale+Hines2006}.

Brian from http://briansimulator.org/ is a generic brain simulator written
in python
\citep{Goodman+Brette2008}.
\subsubsection{astroweb.m}
astroweb.m from http://neuralengr.com/public/uploads/astroweb.m is a Matlab
code. This is a very simple model and code which also runs under Octave,
and is easily rewritten into C, that provides a simulation of the
influence of increased $ \mathrm{Ca}^{2+} $ in astrocytes on release of
glutamate excitatory transmitters from astrocytes, which can participate
in the total transmission of chemical quanta across synaptic gaps when
neurons fire
\citep{Reato+Cammarota+Parra+Carmignoto2012}. This model is geared to
study seizure generation by providing excitatory feedback on neurons
already closed to ``threshold''. Neuronal firings are not simulated in
this model, and the $ \mathrm{Ca}^{2+} $ ions considered do not arise
arise from cooperative regenerative processes from internal stores that
are a major determinant of $ \mathrm{Ca}^{2+} $ waves
\citep{Ross2012}. The author of this simulation admits: ``While the model
needs to be improved, it was a first attempt to give a simple
description of neuron-astrocyte interaction suitable for large scale
network simulations.'' This also is a reasonable appraisal of the state
of research and knowledge of multiple-scale and large-scale neocortical
interactions.
\subsubsection{2-D model of $ \mathrm{Ca}^{2+} $ waves in glia}
At another limit of current modeling, a relatively complex model of
neuron-glia interactions interactions that includes intracellular $
\mathrm{Ca}^{2+} $ wave (ICW) spread in glia is developed as a system of
stochastic differential equations
\citep{Iacobas+Suadicani+Spray+Scemes2006}. This ICW model includes
realistic complexity of chemical and biological interactions, but it has
not yet been developed to the stage of including large-scale many
glial-neuron interactions.
\subsubsection{Other neuron-astrocyte models}
There are some other neuron and neuron-astrocyte models that we continue
to study for elements that may be useful to our own simulation
\citep{Amiri+Montaseri+Bahrami2013,Gerstner+Brette2009,Kang+Othmer2009,Liu+Li2013,Shigetomi+Bushong+Haustein+Tong+Jackson-Weaver+Kracun+Xu+Sofroniew+Ellisman+Khakh2013}.
We need to explicitly introduce $ \mathrm{Ca}^{2+} $ wave dynamics into
tripartite interactions, to further study the influence of EEG $ \mathbf
{A} $ on $ \mathrm{Ca}^{2+} $ wave momenta $ \mathbf{p} $.
\clearpage
\subsubsection{Discussion}
Research to date clearly supports the importance of tripartite synaptic
processes influenced by $ \mathrm{Ca}^{2+} $ waves. Additional
experimental and theoretical development will make it possible to
simulate the influence of these interactions at the larger scale of
columnar interactions, e.g., as $ \mathrm{Ca}^{2+} $ waves influence
averaged synaptic processes in columnar interactions.
\subsection{Quantum coherence simulations}
We are researching the calculation of extended quantum coherence of $
\mathrm{Ca}^{2+} $ using PATHINT
\citep{Ingber+Nunez1995} or PATHTREE
\citep{Ingber+Chen+Mondescu+Muzzall+Renedo2001} code used in previous
publications by the author. The use of these codes for path-integral
calculations, in contrast to Monte Carlo codes, permits a time step-wise
propagation of quite general time-dependent nonlinear multivariate
propagators, during which new events may enter the propagation, e.g.,
simulating BB decoupling from interacting ions in $ \mathrm{Ca}^{2+} $
waves, promoting long coherence times. Path integral methods have been
used for other biophysical systems
\citep{Nalbach+Ishizaki+Fleming+Thorwart2011}. This is quite a familiar
situation in financial derivatives like options, and these codes were
used successfully in that discipline, e.g., much better than using Monte
Carlo calculations of path integrals
\citep{Ingber+Chen+Mondescu+Muzzall+Renedo2001}. However, here the
propagator may not be easily defined (we have not found any good example
yet), and it likely will live in complex $ (x+\mathbf{\hat{i}}y) $ space
which makes numerical details quite harder.

Quite apart from obvious possible extrapolations to the realm of
interactions between macroscopic processes involved in consciousness and
the quantum scales of neocortical processes, the issues in this project
are within currently accessible experimental and theoretical physics per
se.
\clearpage
\subsubsection{Discussion}
Experimental and theoretical research to date does not dismiss the
importance of quantum molecular processes directly influencing larger
scale interactions. On the contrary, a case has been made that coupling
of molecular processes for extended coherence times at the level of $
\mathrm{Ca}^{2+} $ waves interacting with large-scale processes at the
level of scalp EEG, via $ \mathbf{\Pi }=\mathbf{p}+q\mathbf{A} $, should
continue to be investigated.
\section{Conclusion}
A model has been developed to calculate and experimentally test the
coupling of molecular scales of $ \mathrm{Ca}^{2+} $ wave dynamics with $
\mathbf{A} $ fields developed at macroscopic regional scales measured by
coherent neuronal firing activity measured by scalp EEG\@.

For several decades biological and biophysical research into neocortical
information processing has explained neocortical interactions as
specific bottom-up molecular and smaller-scale processes
\citep{Rabinovich+Varona+Selverston+Arbaranel2006}. It is clear that most
molecular approaches consider it inevitable that their approaches at
molecular and possibly even quantum scales will yet prove to be causal
explanations of relatively macroscopic phenomena.

This study crosses molecular, microscopic (synaptic and neuronal),
mesoscopic (minicolumns and macrocolumns), and macroscopic regional
scales. Over the past three decades, with regard to STM and LTM
phenomena, which themselves are likely components of other phenomena
like attention and consciousness, the SMNI approach has yielded specific
details of STM not present in molecular approaches
\citep{Ingber2012a}. The SMNI calculations detail information processing
capable of neocortex using patterns of columnar firings, e.g., as
observed in scalp EEG
\citep{Salazar+Dotson+Bressler+Gray2012}, which give rise to a SMNI
vector potential $ \mathbf{A} $ that influences the molecular $ \mathrm{Ca}^
{2+} $ momentum $ \mathbf{p} $, and thereby synaptic interactions.
Explicit Lagrangians have been given, serving as cost/objective
functions that can be fit to EEG data, as similarly performed in
previous SMNI papers.

Considerations in both classical and quantum physics predict a
predominance of $ \mathrm{Ca}^{2+} $ waves in directions closely aligned
to the direction perpendicular to neocortical laminae ($ \mathbf{A} $ is
in the same direction as the current flow, typically across laminae,
albeit they are convoluted), especially during strong collective EEG (e.g.,
strong enough to be measured on the scalp, such as during selective
attention tasks). Since the spatial scales of $ \mathrm{Ca}^{2+} $ wave
and macro-EEG are quite disparate, an experimenter would have to be able
to correlate both scales in time scales on the order of tens of
milliseconds.

The basic premise of this study is robust against much theoretical
modeling, as experimental data is used wherever possible for both $
\mathrm{Ca}^{2+} $ ions and for large-scale electromagnetic activity.
The theoretical construct of the canonical momentum $ \mathbf{\Pi }=\mathbf
{p}+q\mathbf{A} $ is firmly entrenched in classical and quantum
mechanics. Calculations demonstrate that macroscopic EEG $ \mathbf{A} $
can be quite influential on the momentum $ \mathbf{p} $ of $ \mathrm{Ca}^
{2+} $ ions, at scales of both classical and quantum physics.

A single $ \mathrm{Ca}^{2+} $ ion can have a momentum appreciably
altered in the presence of macrocolumnar EEG firings, and this effect is
magnified when many ions in a wave are similarly affected. Therefore,
large-scale top-down neocortical processing giving rise to measurable
scalp EEG can directly influence molecular-scale bottom-up processes.
This suggests that, instead of the common assumption that $ \mathrm{Ca}^
{2+} $ waves contribute to neuronal activity, they may in fact at times
be caused by the influence of $ \mathbf{A} $ of larger-scale EEG\@. The
SMNI model supports a mechanism wherein the $ \mathbf{p}+q\mathbf{A} $
interaction at tripartite synapses, via a dynamic centering mechanism (DCM)
to control background synaptic activity, acts to maintain STM during
states of selective attention. Such a top-down effect awaits forensic in
vivo experimental verification, requiring appreciating the necessity and
due diligence of including true multiple-scale interactions across
orders of magnitude in the complex neocortical environment.
\section*{Acknowledgment}
We thank the National Science Foundation Extreme Science and Engineering
Discovery Environment (XSEDE.org), for grant PHY130022,
``Electroencephalographic field influence on calcium momentum waves''.
Lester Ingber thanks Paul Nunez and William Ross for verification of
some experimental data, Charlie Gray for a preprint, and Danko Georgiev
and Davide Reato for helpful discussions.
\clearpage
\section*{References}
\bibliographystyle{elsarticle-harv.bst}
\bibliography{lingber}
\ %
\\
\textdollar{}Id:  http://ingber.com/smni14\_eeg\_ca.pdf 1.380 2013/11/23
19:39:16 ingber Exp ingber\textdollar{}
\end{document}